\documentclass[10pt,journal,compsoc]{IEEEtran}
\usepackage{url}
\usepackage{amsfonts}
\usepackage{amsmath}
\usepackage{amssymb}
\usepackage{color}
\usepackage{graphicx} 
\usepackage{epstopdf}
\usepackage{bm}
  
\usepackage{dcolumn}
\usepackage{bm}
\usepackage{hyperref}
\usepackage{dsfont}
\usepackage[caption=false]{subfig}
\usepackage{float}
\usepackage{microtype}
\newcommand{\Is}{\textls[-50]{Is}}
\newcommand{\Ia}{\textls[-50]{Ia}}
\newtheorem{theorem}{Theorem}

\begin{document}

\title{A Non-Markovian Model to Assess Contact Tracing for the Containment of COVID-19}

\author{Aram Vajdi$^{*}$ , Lee W. Cohnstaedt$^{1,\ddagger}$,  Leela E. Noronha$^{\ddagger}$, Dana N. Mitzel$^{\ddagger}$, William C. Wilson$^{\ddagger}$, and Caterina M. Scoglio$^{*}$

 \IEEEcompsocitemizethanks{\IEEEcompsocthanksitem $^{1}$ Corresponding author 
\protect\\
$^{*}$ Department of Electrical and Computer Engineering, Kansas State University,
Manhattan, KS 
\protect\\ $^{\ddagger}$ United States Department of Agriculture, Agricultural Research Service, Foreign Arthropod-Borne Animal Diseases Research Unit, Manhattan, Kansas
}     
}

\IEEEtitleabstractindextext{%
\begin{abstract}
COVID-19 remains a challenging global threat with ongoing waves of infections and clinical disease which have resulted millions of deaths and an enormous strain on health systems worldwide. Effective vaccines have been developed for the SARS-CoV-2 virus and administered to billions of people; however, the virus continues to circulate and evolve into new variants for which vaccines may ultimately be less effective. Non-pharmaceutical interventions, such as social distancing, wearing face coverings, and contact tracing, remain important tools, especially at the onset of an outbreak. In this paper, we assess the effectiveness of contact tracing using a non-Markovian, network-based mathematical model. To improve the reliability of the novel model, empirically determined distributions were incorporated for the transition time of model state pairs, such as from exposed to infected states. The first-order closure approximation was used to derive an expression for the epidemic threshold, which is dependent on the number of close contacts. Using survey contact data collected  during the 2020 fall academic semester from a university population, we determined that even four to five contacts are sufficient to maintain viral transmission. Additionally, our model reveals that contact tracing can be an effective outbreak mitigation measure by reducing the epidemic size by more than three-fold. Increasing the reliability of epidemic models is critical for accurate public health planning and use as decision support tools. Moving toward more accurate non-Markovian models built upon empirical data is important.
\end{abstract}

}

\maketitle

\IEEEdisplaynontitleabstractindextext

%
\IEEEpeerreviewmaketitle

\IEEEraisesectionheading{\section{Introduction}\label{sec:introduction}}
\IEEEPARstart{C}{ontact}
%
%
%
%
tracing is a primary public health response to infectious disease outbreaks \cite{chowdhury2020covid,kretzschmar2020impact}. The adoption of contact tracing to contain COVID-19 met considerable challenges, due to the intensive process related to manual contact tracing and the hesitancy for adopting app-based contact tracing tools\cite{barrat2020effect}. Assessing the impact of contact tracing as a control measure remain of great importance.

In \cite{keeling2020efficacy}, the authors derived contact patterns in the UK using a postal and online cross-sectional survey. They predicted that, under effective contact tracing, fewer than one in six cases would generate any subsequent infections that will not be traced. This comes at a high cost, with an average of 36 individuals traced per infected case. Making the definition of a close contact more stringent can reduce this cost, but with increased risk of untraced cases.

A mathematical model assessing contact tracing and isolation for COVID-19 has been developed in \cite{kucharski2020effectiveness}. The study estimated that a high proportion of cases would need to self-isolate and a high proportion of their contacts successfully traced to contain the epidemic. If combined with moderate physical distancing measures, self-isolation and contact tracing would likely achieve control of SARS-CoV-2 transmission.

Another aspect of COVID-19 that can alter the effectiveness of contact tracing is the high transmissibility of SARS-CoV-2 before and immediately after symptom onset \cite{cheng2020contact}. Therefore, finding and isolating symptomatic patients alone may not suffice to interrupt transmission, and that more generalized measures might be required, such as social distancing.

Considering the act of quarantining susceptible individuals (uninfected contacts) as the contact tracing cost, a model developed in \cite{moon2021contact} has shed light on an interesting phenomenon. The number of quarantined susceptible people increases with the increase of tracing because each confirmed case increases the number of total contacts. However, there is an inflection point after which the number of traces decreases with increased tracing because there are fewer confirmed cases. 

Models for assessing contact tracing have been developed for many infectious diseases, before the COVID-19 pandemic. A systematic review of these models can be found in \cite{kwok2019epidemic} which, reviewed mathematical models for contact tracing and follow-up control measures of SARS and MERS transmission. All models required data to estimate specific model parameter distributions. A major concern identified was whether public health administrators can collect all the required data for building epidemiological models in a short period of time during the early phase of an outbreak where contact tracing is more effective.

The analysis of case data from the Wuhan COVID-19 outbreak highlighted non-exponential distributions for critical transition times during infection, such as the infectious period and the incubation period \cite{sanche2020high,backer2020incubation}. Simulations show that different distributions of the infectious period duration with the same mean values lead to different epidemic curves.

Consequently, it is critical to develop models that can input empirical distributions, often non-exponential, for the transition time between compartments.

The majority of network-based individual level pathogen spread models assume exponential distributions for transition times among state pairs. However, these models need adaptation to incorporate general transition time distributions. One possible approach is based on the adoption of phase-type distributions \cite{nowzari2015general}.

A phase-type distribution can mathematically represent the transition between any two states in a pathogen spread model. By  expanding the source state of the original model into a series of $q$ states, then any arbitrary distribution can be approximated as a phase-type distribution to represent the time distribution between subsequent states. Techniques exist to estimate the new parameters to realize the desired distribution. When the transition time distribution between the two original states is general, this distribution can be approximated by transitions between subsequent pairs of states among the q states that are exponentially distributed. When the objective is to derive analytical results concerning non-Markovian models, both first-order \cite{sherborne2018mean} and second order \cite{pellis2015exact} closure techniques can be used. For efficient methods to simulate non-Markovian models Gillespie based algorithms are available \cite{boguna2014simulating}.

The goal of this paper is to develop a non-Markovian model, both exact and first-order closure approximation, appropriate for assessing contact tracing and to determine the epidemic threshold. A new networked compartmental model with compartments able to represent the disease evolution and the contact tracing process is presented in section \ref{model description}. The model is general to allow arbitrary transition time distributions and the mean field equations for such a non-Markovian model is derived. A mathematical analysis was performed to derive the epidemic threshold, in section \ref{analysis}. The model was applied to the Kansas State University case study reported in section \ref{Case study}.
Summarizing, the original contributions of the paper are the following:
\begin{itemize}
\item	A novel and general network-based model structure for pathogen transmission, 
\item	The determination of the first-order epidemic threshold, 
\item	Numerical results about the effectiveness of contact tracing in the case study of Kansas State University.
\end{itemize}
\section{Model description}\label{model description} 
We begin by presenting a continuous time non-Markovian transmission model to describe the tractability of COVID-19 spread when contact tracing strategies are implemented in a population. 
\begin{figure}[t]
\centering
   \includegraphics[width=1\columnwidth]{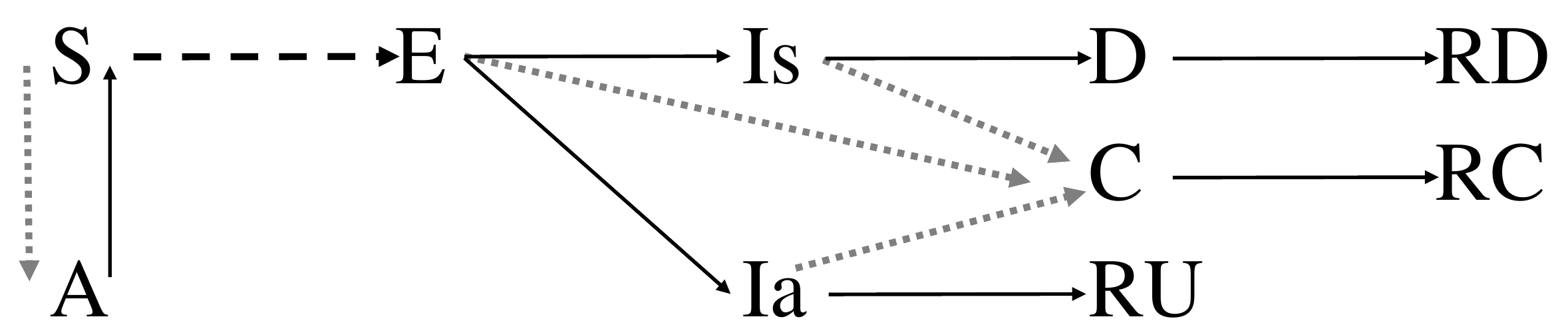} %
   \caption{Diagram of transitions in the SAIDR spreading model. An individual can be in one of the following 10 states: susceptible (S), susceptible alert (A), infected but not not infectious (E), infectious presymptomatic or symptomatic (Is), infectious asymptomatic (Ia), detected symptomatic cases (D) that trigger contact tracing, confirmed cases (C) through contact tracing, and removed states (RD), (RU), (RC) that do not cause new infection or contact tracing. Arrow type in the diagram represent the transitions mechanism. Dotted and dashed arrows show transitions that are due to the individual's contacts and solid arrows denote spontaneous transitions that are not forced by interactions.}   
\label{trans1}
\end{figure}
The model assumes individuals (nodes) are in one of ten states. If individuals are susceptible (S), they can become infected but not immediately infectious, i.e,  exposed (E), through interaction with infectious individuals that are either symptomatic (Is) or asymptomatic (Ia). The model assumes that the infectious asymptomatic state does not contain the presymptomatic cases that present symptoms at some point during the course of infection. Presymptomatic and symptomatic states are combined into the infectious symptomatic (Is) state. Furthermore, the model assumes the fraction of individuals in the E state that moves to the Is state is $p_{\Is}$, and an asymptomatic infectious (Ia) individual cannot be detected unless tested. Therefore, an individual in the Ia state continues spreading virus while infectious. The model assumes that a person in the Ia state is only infectious for a period of time and subsequently move to the removed state RU.

Symptomatic individuals present symptoms and therefore transition to detected state (D) even if not tested. This triggers contact tracing among the contacts.  The assumption in this model is some contacts from detected cases get tested, and if they are infected, they move to confirmed traced state (C), and if virus negative, they transition to alert traced (A) state for a period of time. Indeed, testing capacity could be limited, and the contact tracing strategy may only require quarantining of close contacts. In such a case, we can regard the alert state as quarantine or a subset of S but temporarily removed from the susceptible population. Moreover, we can add more states to the model and divide the confirmed (C) state into three sub-states, to differentiate the disease state of the traced contacts. However, to keep the model simple, the model only considers one C state. Since an individual induces contact tracing as long as is in the D state, after a period of time it should moves to the removed (RD) state so that the contact tracing stops. Similarly, the model assumes an individual in the C state initiates contact tracing and later moves to removed state (RC). 
We refer to the transmission model described above as SAIDR, and Fig. \ref{trans1} shows the diagram of the transitions that a person may experience in the SAIDR model. Transition from the S state to the E state is induced by contacts with individuals in the Is or Ia states. Transitions shown by dotted arrows are due to contact tracing and are induced by contacts with individuals in the C or D states. If a transition is not induced by any interaction, we call it nodal transition and it is shown by solid arrow. 

In the SAIDR model, we assume the transition times are random variables drawn from appropriate distributions for the state. The traditional approach in the analysis of
epidemic models assumes the distribution of transition times to be exponential, but this might not be an accurate description of all the processes considered in our spreading model. For instance, real-world data suggest transition times for the E $\rightarrow$ I process are not distributed exponentially. Some factors that have been suggested to potentially influence these transition times, specifically the incubation period (E → Is), include age, infectious dose, and physiological stressors. 

Moreover, allowing non-exponential distributions for auxiliary processes such as D $\rightarrow$ RD, increases the degrees of freedom in the model. Hence, here we allow the nodal transitions to have non-exponential distributions. An exponential distribution is specified by the rate parameter $\lambda$, and the assumption that a transition time is exponentially distributed implies for any infinitesimal period of time $dt$ the transition happens with a constant probability $\lambda\ dt$, regardless of the age of the process. Indeed, such a constant rate of transition is a suitable approximation for the transmission process. Therefore, for the transmission process S $\rightarrow$ E, the model assumes the transition times are exponentially distributed as long as the infecting individual stays infected. Additionally, if a susceptible person has several infectious contacts, the model assumes the infecting processes are independent. Similarly, in the contact tracing modeling, we use exponential distribution for the transition times of processes that change state of an individual to the C or A states.
 
Finally, to model the interactions that result in virus transmission or lead to contact tracing, we use an undirected network where the nodes represent individuals in the population and the links denote the contacts among them. Consider a network $G = G(V,E,W_{inf},W_{tr})$, with $V$ representing the set of nodes, $E$ a set of links between the nodes and $W_{inf},W_{tr} : E \rightarrow  [0\  1]$ weight functions defined over the links. We use the weight functions $W_{inf},W_{tr}$ to quantify heterogeneity of the contacts in relation to virus transmission probability or contact tracing probability, respectively. In other words, for each contact, we modify the rate of exponential distributions for the infecting and contact tracing processes multiplying the rates by the contact weights. This allows us to define various types of contacts with different probabilities for virus transmission or contact tracing.

\section{Mathematical analysis of the model}\label{analysis}
In this section we develop a set differential equations that approximately describes the behavior of the transmission model discussed in section \ref{model description}. Indeed, there is an extensive body of research concerning Markovian spreading unfolding over networks \cite{van2011n,sahneh2013generalized,pastor2015epidemic,newman2002spread,boguna2003absence,chakrabarti2008epidemic,goltsev2012localization}, and it is shown the exact mathematical treatment of such a system is not tractable because we need to follow the joint state of all the nodes. Therefore, a mean-field type approximation, which assumes statistical independence of neighboring nodes' states, is often employed to study network spreading models. Here, we limit the non-exponential random variables in the spreading model of \ref{model description} to the Erlang family of distributions. This enables us to use a characteristic of Erlang distribution to cast our spreading model into a Markovian model, for which we can develop an N-intertwined set of differential equations \cite{van2011n}. 
\begin{figure*}[t]
\centering
   \includegraphics[width=2\columnwidth]{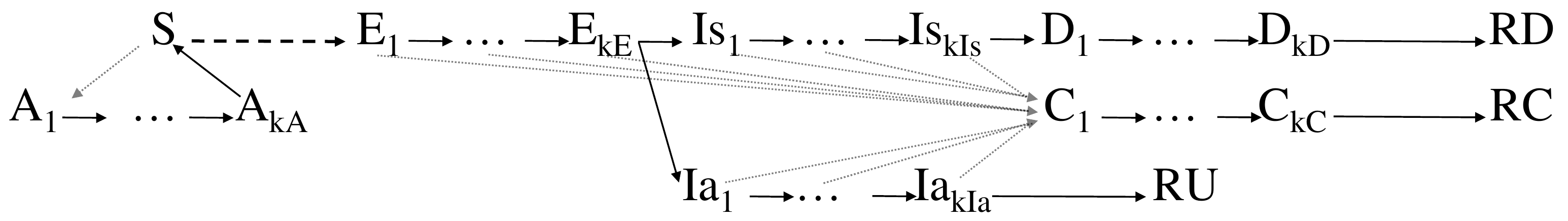} %
   \caption{Diagram of the SAIDR spreading model when the A$\rightarrow$S, E$\rightarrow$I, Is$\rightarrow$D, Ia$\rightarrow$RU, D$\rightarrow$RD and C$\rightarrow$RC transitions are replaced with successive auxiliary Markovian transitions. This leads to actual transition times with Erlang distributions.}
\label{trans2}
\end{figure*}
The Erlang distribution is a two-parameter family of continuous probability distributions with the density function 
\begin{equation}\label{erlang}
f(t|k,\lambda)=\frac{\lambda^{k}t^{k-1}e^{-\lambda t}}{(k-1)!} \ \ \ \ \   \text {for} \ t \in \left[ 0 \ \infty\right), 
\end{equation}
where the shape parameter $k$ is a positive integer, and $\lambda>0$ is called the rate parameter. From equation above, we can see that the exponential distribution is a special case of Erlang distribution with $k=1$, and the mean and variance of Erlang distribution are $k/\lambda$ and $k/\lambda^{2}$, respectively. Since the Erlang distribution has two parameters, we can adjust them to obtain a density function with a desired mean and a variance close to a target value. Thus, we can use the Erlang distribution as an approximation for a large class of distributions. 

One important characteristic of the Erlang distribution can be stated as follows: distribution of sum of $k$ independent random variables, each having an exponential distribution with rate $\lambda$, is Erlang distribution with the shape parameter $k$ and the rate parameter $\lambda$, i.e.
\[
\text{if }t_{i} \thicksim Exp(\lambda), \text{ then } t=\sum_{i}^{k}t_{i}\thicksim Erlang(k,\lambda).
\]

This characteristic implies if the random transition time of a process follows Erlang($k,\lambda$) distribution, the process can be modeled as $k$ successive transitions between $k+1$ auxiliary states where each transition time is exponentially distributed with the rate $\lambda$. Therefore, if a nodal transition time in the spreading model of Fig. \ref{trans1} has an Erlang type distribution, it is possible to replace the nodal transition with a set of successive Markovian transitions and obtain an equivalent spreading model. Fig. \ref{trans2} shows a spreading model where all the transition times are exponentially distributed and it is equivalent of the original model in Fig. \ref{trans1}. Indeed, even if the distribution of a transition time is not Erlang, we still can use phase-type distribution method to obtain  a mixture of Markovian transitions that is approximately equivalent to the original transition \cite{nowzari2015general}. However, we limit the distribution of nodal transition times to Erlang distribution because it incorporates a family of density functions with arbitrary means and variances.

Considering the Markovian spreading model in Fig. \ref{trans2}, let $s^{i}$ represents state of node $i$ among $M$ possible states in the model, and $\mathbb{S}=[s^{1},s^{2},\cdots, s^{N}]$ represents the joint state of all the individuals in the population. Based on the nodal description of the processes, we deduce the joint state is a continuous-time Markov chain over a space consisting of $M^{N}$ possible network states. Therefore, the exact mathematical treatment of the system is not tractable when the number of nodes is large. But if we use mean-field approximation \cite{van2011n}, that assumes statistical independence of neighboring nodes' states, we obtain a set of $M\times N$ equations, which we can solve computationally. This approximation assumes at any time, the joint probability of finding nodes $i$ and $j$ in the states $s^{i}$ and $s^{i}$ can be written as the multiplication of marginal probabilities, $\Pr(s^{i})\times \Pr(s^{j})$.

To formulate the approximate behavior of the Markovian model depicted in Fig. \ref{trans2}, we use dynamic variables, $S^{i}(t),A^{i}_{1}(t),\cdots,A^{i}_{k_{A}}(t),E^{i}_{1}(t),\cdots,E^{i}_{k_{E}}(t),\cdots,RU^{i}(t) $, which represent the probabilities of finding node $i$ in the corresponding states. If $r^{i}_{I}$ and $r^{i}_{T}$ denote the rates for the transmission and contact tracing  processes exerted on node $i$, we have
\begin{equation}\label{forces}
\begin{split}
&r^{i}_{I}= \sum_{j}W_{1}^{i,j} \left(\beta_{\Is}\sum_{k} \Is ^{j}_{k}+\beta_{\Ia}\sum_{k}\Ia^{j}_{k} \right),\\
&r^{i}_{T}= \sum_{j}W_{2}^{i,j} \lambda_{tr} \left(\sum_{k}C^{j}_{k}+\sum_{k}D^{j}_{k} \right),
\end{split}
\end{equation}
where $\lambda_{tr}$ is the rate for the the contact tracing process, and we have assumed symptomatic and asymptomatic infectious states have different infectiousness which is reflected in the infectious rates parameters, $\beta_{\Is}$, $\beta_{\Ia}$.  
Finally, using the joint state independence approximation, we can write the following set of equations for all the nodes in the network

\begin{equation}\label{dynamics}
\begin{split}
&\dot{S}^{i}(t)=-r^{i}_{I}S^i-r^{i}_{T}S^i+A^i_{k_A}\lambda_A\\
&\dot{A}_1^{i}(t)=r^{i}_{T}S^i-\lambda_A A^i_{1}\\
&\dot{A}_k^{i}(t)=\lambda_A \left(A^i_{k-1}-A^i_{k}\right)\ \  \text{for}\ \ k=2,\cdots,k_A\\
&\dot{E}_1^{i}(t)=r^{i}_{I}S^i-\lambda_E E^i_{1}-r^{i}_{T}E_1^{i}\\
&\dot{E}_k^{i}(t)=\lambda_E \left(E^i_{k-1}-E^i_{k}\right)-r^{i}_{T}E_k^{i}\ \  \text{for}\ \ k=2,\cdots,k_E \\
&\dot{\Is}_1^{i}(t)=p_{\Is}\lambda_E E_{k_E}^i-\lambda_{\Is} {\Is}^i_{1}-r^{i}_{T} {\Is}_1^{i}\\
&\dot{\Is}_k^{i}(t)=\lambda_{\Is} \left({\Is}^i_{k-1}-{\Is}^i_{k}\right)-r^{i}_{T}{\Is}_k^{i}\ \  \text{for}\ \ k=2,\cdots,k_{\Is}\\
&\dot{\Ia}_1^{i}(t)=p_{\Ia}\lambda_E E_{k_E}^i-\lambda_{\Ia} {\Ia}^i_{1}-r^{i}_{T} {\Ia}_1^{i}\\
&\dot{\Ia}_k^{i}(t)=\lambda_{\Ia} \left({\Ia}^i_{k-1}-{\Ia}^i_{k}\right)-r^{i}_{T}{\Ia}_k^{i}\ \  \text{for}\ \ k=2,\cdots,k_{\Ia}\\
&\dot{D}_1^i(t)=\lambda_{\Is}{\Is}^i_{k_{\Is}}-\lambda_D D^i_1\\
&\dot{D}_k^i(t)=\lambda_D \left( D^i_{k-1}-D^i_{k} \right) \ \  \text{for}\ \ k=2,\cdots,k_{D}\\
&\dot{C}_1^i(t)=r^{i}_{T} \left( \sum_k E_k^i+\sum_k {\Is}^i_{k} +\sum_k {\Ia}^i_{k} \right) - \lambda_C C^i_1\\\
&\dot{C}_k^i(t)=\lambda_C \left( C^i_{k-1}-C^i_{k} \right) \ \  \text{for}\ \ k=2,\cdots,k_{C}\\
&\dot{RC}^i(t)=\lambda_C C^i_{k_C}\\
&\dot{RD}^i(t)=\lambda_D D^i_{k_D}\\
&\dot{RU}^i(t)=\lambda_{\Ia}  {\Ia}^i_{k_{\Ia}}
\end{split}
\end{equation}
The system of nonlinear ordinary differential equations
(ODEs) above, describes evolution of nodal states' probability. We can use this set of equations to study behavior of the spreading processes or to estimate the model parameters. In addition, stability analysis of disease-free state of the system leads to derivation of epidemic threshold which establishes a condition that determines
whether an initial infected population will vanish or has the possibility to grow.  

To derive the epidemic threshold the first step is to linearize the nonlinear system \ref{dynamics} about the disease-free steady state. In the disease-free state 
the probability of finding nodes in any state other than the susceptible state is zero. After linearization we arrive at an independent subsystem that describes production of new infections. Since the variables in this subsystem drive other variables in the larger system, stability of this subsystem determines stability of the larger system. This subsystem can be written as follows:
\begin{equation}\label{ldynamics2}
\begin{split}
&\dot{E}_1^{i}(t)=r^{i}_{I}-\lambda_E E^i_{1}\\
&\dot{E}_k^{i}(t)=\lambda_E \left(E^i_{k-1}-E^i_{k}\right)\ \  \text{for}\ \ k=2,\cdots,k_E \\
&\dot{\Is}_1^{i}(t)=p_{\Is}\lambda_E E_{k_E}^i-\lambda_{\Is} {\Is}^i_{1}\\
&\dot{\Is}_k^{i}(t)=\lambda_{\Is} \left({\Is}^i_{k-1}-{\Is}^i_{k}\right)\ \  \text{for}\ \ k=2,\cdots,k_{\Is}\\
&\dot{\Ia}_1^{i}(t)=p_{\Ia}\lambda_E E_{k_E}^i-\lambda_{\Ia} {\Ia}^i_{1}\\
&\dot{\Ia}_k^{i}(t)=\lambda_{\Ia} \left({\Ia}^i_{k-1}-{\Ia}^i_{k}\right)\ \  \text{for}\ \ k=2,\cdots,k_{\Ia}.
\end{split}
\end{equation}

\begin{theorem}\label{th1}
Assuming the infection weight matrix $\mathbf{W_{inf}}$ is irreducible, the disease-free steady state is a locally stable fixed
point of the dynamical system \ref{dynamics} if the following condition holds:
\begin{equation}\label{threshold}
\left(\beta_{\Is} p_{\Is}k_{\Is}\lambda_{\Is}^{-1}+ \beta_{\Ia} p_{\Ia}k_{\Ia}\lambda_{\Ia}^{-1}\right)\rho(\mathbf{W_{inf}})<1,
\end{equation} 
where $\rho(\mathbf{W_{inf}})$ denotes the spectral radius of the weight matrix $\mathbf{W_{inf}}$.
\end{theorem}

To derive the threshold condition in equation \ref{threshold}, we rewrite the linearized subsystem \ref{ldynamics2} in a matrix form where the infection processes, represented by a matrix $\mathbf{\Delta}$, are separated from other transitions,

\begin{equation}\label{lmat}
\begin{split}
&\frac{d\mathbf{X}}{dt}=\left(\mathbf{\Delta}+\mathbf{\Sigma}\right)\mathbf{X}\\
&\mathbf{X}=\left[\mathbf{x}^1,\cdots,\mathbf{x}^N\right]^\top\\
&\mathbf{x}^i=\left[ E_1^{i},\cdots,E_{k_E}^i,{\Is}^i_{1},\cdots,{\Is}^i_{k_{\Is}},{\Ia}^i_{1},\cdots,{\Ia}^i_{k_{\Ia}} \right] 
\end{split}
\end{equation}
The square matrix $\mathbf{\Delta}$ can be written as the   Kronecker product of the network weight matrix $\mathbf{W_{inf}}$ and a matrix $\boldsymbol{\delta}$ which represents the rates and the driving nodal states in the infection process, 
\begin{equation}\label{delmat}
\begin{split}
&\mathbf{\Delta}=\mathbf{W_{inf}}\otimes\boldsymbol{\delta}=\begin{pmatrix}
W^{1,1}_{inf}\boldsymbol{\delta} & \cdots & W^{1,N}_{inf}\boldsymbol{\delta} \\
\vdots  & \ddots & \vdots  \\
W^{N,1}_{inf}\boldsymbol{\delta} &\cdots & W^{N,N}_{inf}\boldsymbol{\delta} 
\end{pmatrix}\\
&\boldsymbol{\delta} = 
\begin{pmatrix}
 \mathbf{0}_{1,k_E}&\beta_{\Is} \mathbf{J}_{1,k_{\Is}},&\beta_{\Ia} \mathbf{J}_{1,k_{\Ia}}\\
  \mathbf{0}_{k_E+k_{\Is}+k_{\Ia}-1,k_E}&\mathbf{0}_{k_E+k_{\Is}+k_{\Ia}-1,k_{\Is}}&\mathbf{0}_{k_E+k_{\Is}+k_{\Ia}-1,k_{\Ia}}
\end{pmatrix}
\end{split}
\end{equation}
where $\mathbf{0}_{k_1,k_2}$ and $\mathbf{J}_{k_1,k_2}$ are $k_1 \times k_2$ matrices of zeros and ones, respectively, 
\begin{equation}\label{jf}
\mathbf{0}_{k_1,k_2} = 
\begin{pmatrix}
 0&\cdots\\
 \vdots&\ddots
\end{pmatrix}_{k_1\times k_2},
\ \  \mathbf{J}_{k_1,k_2} = 
\begin{pmatrix}
 1&\cdots\\
 \vdots&\ddots
\end{pmatrix}_{k_1\times k_2}.
\end{equation}
The matrix $\mathbf{\Sigma}$ in equation \ref{lmat} is a block diagonal matrix which can be written as the Kronecker product of the identity matrix of size $N$ and a matrix $\boldsymbol{\sigma}$ which represents nodal transitions,
\begin{equation}\label{smat}
\begin{split}
&\mathbf{\Sigma}=\mathbf{I}_N\otimes\boldsymbol{\sigma} =\begin{pmatrix}
\boldsymbol{\sigma} & 0&\cdots &0\\
0 & \boldsymbol{\sigma}&\cdots&0  \\
\vdots  & \vdots& \ddots & \vdots \\
0  & 0& \cdots & \boldsymbol{\sigma} \\
\end{pmatrix}_{N \times N}\\
&\boldsymbol{\sigma}=\begin{pmatrix}
\lambda_E \mathbf{H}_{k_E} & \mathbf{0}_{k_E,k_{\Is}}&\mathbf{0}_{k_E,k_{\Ia}} \\
p_{\Is} \lambda_{E} \mathbf{G}_{k_{\Is},k_E} &  \lambda_{\Is} \mathbf{H}_{k_{\Is}}& \mathbf{0}_{k_{\Is},k_{\Ia}}\\
p_{\Ia} \lambda_{E} \mathbf{G}_{k_{\Ia},k_E} &\mathbf{0}_{k_{\Ia},k_{\Is}}& \lambda_{\Ia} \mathbf{H}_{k_{\Ia}}
\end{pmatrix},
\end{split}
\end{equation}
and the matrices $\mathbf{H}, \ \mathbf{G}$ used in the definition of $\boldsymbol{\sigma}$ have the following structures  
\[
\mathbf{H}_k=
\begin{pmatrix}
-1& 0&0&\cdots \\
1 & -1&0&\cdots  \\
0 & 1&-1&\ddots  \\
\vdots  & \ddots&  \ddots&  \ddots\\
\end{pmatrix}_{k\times k},
\mathbf{G}_{k_1,k_2}=
\begin{pmatrix}
\cdots & 0&1 \\
\cdots & 0&0 \\
\reflectbox{$\ddots$} &\vdots &\vdots \\
\end{pmatrix}_{k_1\times k_2}.
\]
In the $\mathbf{H}$ matrix, the diagonal elements are $-1$, the lower diagonal elements are $1$ and the rest of elements are zero. In the $\mathbf{G}$ matrix, the element in the upper-right corner is $1$ and the rest of elements are zero.

The stability of steady state of the linear system \ref{lmat} is determined by the spectral bound of  
the square matrix $\mathbf{\Delta}+\mathbf{\Sigma}$, defined as 
\[
s\left(\mathbf{\Delta}+\mathbf{\Sigma}\right)=\sup \left\lbrace  \Re(\alpha) | \alpha \in \eta \left(\mathbf{\Delta}+\mathbf{\Sigma}\right)\right\rbrace,
\]
where $\eta \left(\mathbf{\Delta}+\mathbf{\Sigma}\right)$
denotes the set of eigenvalues of $\mathbf{\Delta}+\mathbf{\Sigma}$. The linear system is exponentially stable if and only if the real parts of the eigenvalues are negative (i.e., if  $s\left(\mathbf{\Delta}+\mathbf{\Sigma}\right)<0$). To find the condition that leads to a negative spectral bound we apply theorem A.1 from reference \cite{diekmann2010construction}, which, for the sake of completeness, we state as the following theorem. 
\begin{theorem}\label{th2}
If $\mathbf{X}$ is a positive matrix and $\mathbf{Y}$ is a positive off-diagonal matrix with $s\left(\mathbf{Y}\right)<0$, then 
\[
\text{sign}\left(s\left(\mathbf{X}+\mathbf{Y}\right)\right)=\text{sign}\left(\rho\left(-\mathbf{X}\mathbf{Y}^{-1} \right)-1\right)
\]
where $\rho(.)$ and sign(.) denote spectral radius and sign functions, respectively.
\end{theorem}
We note that in reference \cite{diekmann2010construction}, positive matrices are defined as non-zero matrices with all entries non-negative; and a matrix is defined as positive off-diagonal if all entries are non-negative except possibly those on the diagonal. 

To apply theorem \ref{th2}, we first investigate if the matrices $\mathbf{\Delta}$ and $\mathbf{\Sigma}$ satisfy the condition stated in the theorem. 
From the definition of $\mathbf{\Delta}$ and $\mathbf{\Sigma}$ in the equations \ref{delmat},\ref{smat}, we can see $\mathbf{\Delta}$ is a positive matrix and $\mathbf{\Sigma}$ is positive off-diagonal. Moreover, since $\mathbf{\Sigma}$ is positive off-diagonal lemma 6.12 in \cite{diekmann2000mathematical} shows that $s\left(\mathbf{\Sigma}\right)<0$ if and only if $\mathbf{\Sigma}$ is invertible and $-\mathbf{\Sigma}^{-1}$ is a positive matrix. Indeed, we can directly calculate $-\mathbf{\Sigma}^{-1}$ as fallows  
\begin{equation}
\begin{split}
&-\mathbf{\Sigma}^{-1}=\mathbf{I}_N\otimes\left(-\boldsymbol{\sigma}^{-1}\right)\\
&-\boldsymbol{\sigma}^{-1}=\begin{pmatrix}
\lambda_E ^{-1} \mathbf{F}_{k_E} & \mathbf{0}_{k_E,k_{\Is}}&\mathbf{0}_{k_E,k_{\Ia}} \\
p_{\Is} \lambda_{\Is}^{-1 }\mathbf{J}_{k_{\Is},k_E} &  \lambda_{\Is}^{-1}\mathbf{F}_{k_{\Is}}& \mathbf{0}_{k_{\Is},k_{\Ia}}\\
p_{\Ia} \lambda_{\Ia}^{-1} \mathbf{J}_{k_{\Ia},k_E} &\mathbf{0}_{k_{\Ia},k_{\Is}}& \lambda_{\Ia}^{-1} \mathbf{F}_{k_{\Ia}}
\end{pmatrix}
\end{split}
\end{equation}
where the matrices $\mathbf{0}$ and $\mathbf{J}$ are defined in the equation \ref{jf}, and $\mathbf{F}_k$ is a lower triangular matrix of size $k$, with non-zero entries equal 1
\begin{equation}
\begin{split}
\mathbf{F}_k=
\begin{pmatrix}
1& 0&0&\cdots \\
1 & 1&0&\cdots  \\
1 & 1&1&\ddots  \\
\vdots  & \vdots&  \vdots&  \ddots\\
\end{pmatrix}_{k\times k}
\end{split}
\end{equation}
It is clear that $-\mathbf{\Sigma}^{-1}$ is a positive matrix. Therefore we can use theorem \ref{th2} to conclude that 
\[
s\left(\mathbf{\Delta}+\mathbf{\Sigma}\right)<0 \ \ \text{if and only if}\ \ \rho\left(-\mathbf{\Delta}\mathbf{\Sigma}^{-1} \right)<1.
\]
Finally to prove theorem \ref{th1} we will obtain an expression for $\rho\left(-\mathbf{\Delta}\mathbf{\Sigma}^{-1} \right)$. Using the properties of the Kronecker product we can write  
\begin{equation}\label{spec}
\begin{split}
\eta\left(-\mathbf{\Delta}\mathbf{\Sigma}^{-1}\right)&=\eta\left(\mathbf{W_{inf}}\otimes\left(-\boldsymbol{\delta}\boldsymbol{\sigma}^{-1}\right)\right)\\&=\left\lbrace \alpha_i \gamma_j \ | \ \alpha_i \in \eta\left(\mathbf{W_{inf}}\right),\ \gamma_j \in \eta\left(-\boldsymbol{\delta}\boldsymbol{\sigma}^{-1}\right)\right\rbrace,
\end{split}
\end{equation}
where $\eta(.)$ denotes set of eigenvalues. After calculating the matrix $\left(-\boldsymbol{\delta}\boldsymbol{\sigma}^{-1}\right)$, we can see only the first row is non-zero
\[
-\boldsymbol{\delta}\boldsymbol{\sigma}^{-1}=\begin{pmatrix}
 \beta_{\Is} p_{\Is}k_{\Is}\lambda_{\Is}^{-1}+ \beta_{\Ia} p_{\Ia}k_{\Ia}\lambda_{\Ia}^{-1}&\cdots&\beta_{\Ia} \lambda_{\Ia}^{-1}\\
 0&\cdots&0\\
 \vdots&\ddots&\vdots\\
 0&\cdots&0
\end{pmatrix}
\]
and therefore the eigenvalues of the matrix 
$-\boldsymbol{\delta}\boldsymbol{\sigma}^{-1}$ are 
$\beta_{\Is} p_{\Is}k_{\Is}\lambda_{\Is}^{-1}+ \beta_{\Ia} p_{\Ia}k_{\Ia}\lambda_{\Ia}^{-1}$ and $0$. In addition, since we assume $\mathbf{W_{inf}}$ is an irreducible non-negative weight matrix, Perron–Frobenius theorem shows $\mathbf{W_{inf}}$ has a positive eigenvalue, denoted by $\rho(\mathbf{W_{inf}})$, and absolute value of any other eigenvalue of $\mathbf{W_{inf}}$ is strictly smaller than $\rho(\mathbf{W_{inf}})$. Combining these results about the eigenvalues of $\mathbf{W_{inf}}$ and $-\boldsymbol{\delta}\boldsymbol{\sigma}^{-1}$ with the equation \ref{spec}, we find that 
\begin{equation}\label{ttres}
\rho\left(-\mathbf{\Delta}\mathbf{\Sigma}^{-1} \right)=\left(\beta_{\Is} p_{\Is}k_{\Is}\lambda_{\Is}^{-1}+ \beta_{\Ia} p_{\Ia}k_{\Ia}\lambda_{\Ia}^{-1}\right)\rho(\mathbf{W_{inf}}),
\end{equation}
which leads to the proof of theorem \ref{th1}.

\begin{figure}[t]
	\centering
	\subfloat[]{
		\includegraphics[width=0.24\textwidth]{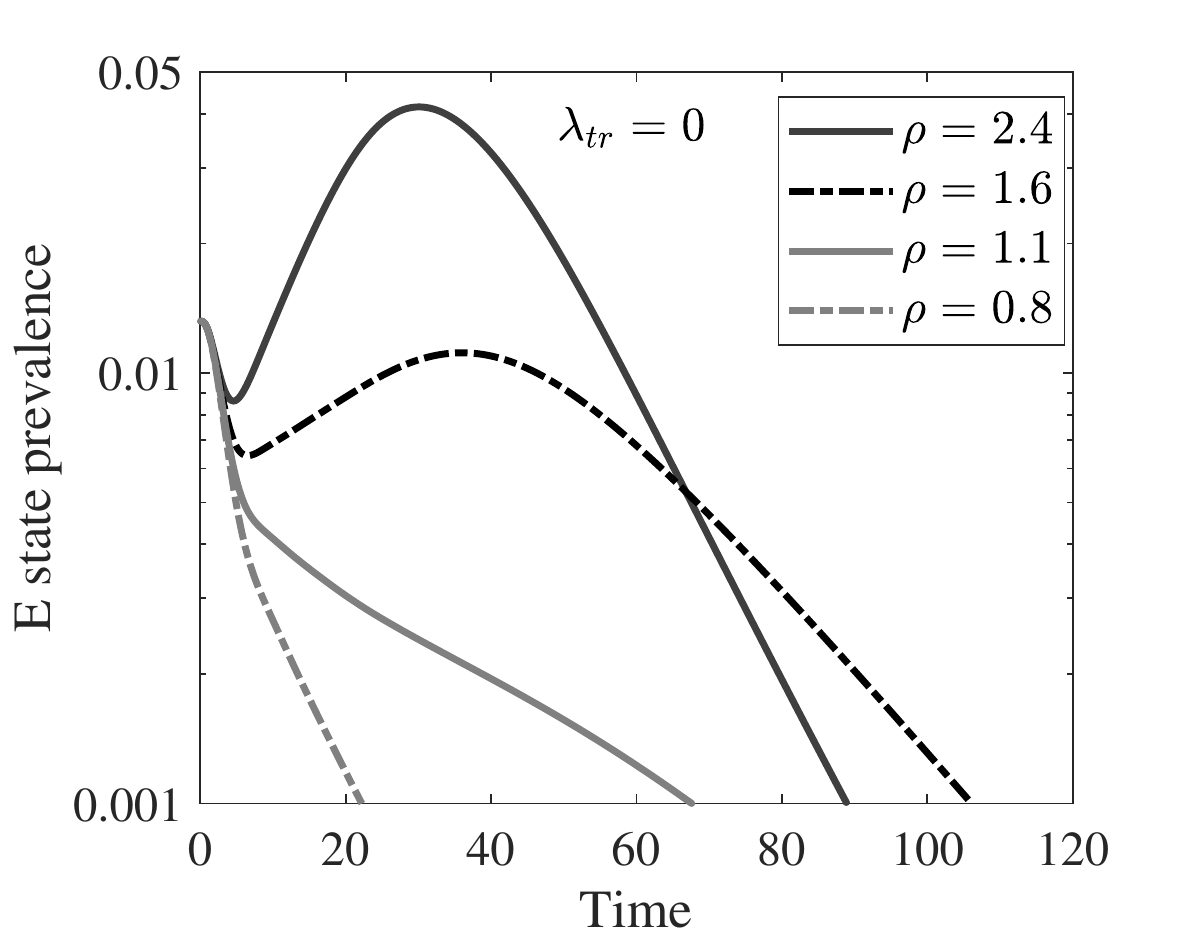}		
		\label{trcon1}} 
	\subfloat[]{
		\includegraphics[width=0.24\textwidth]{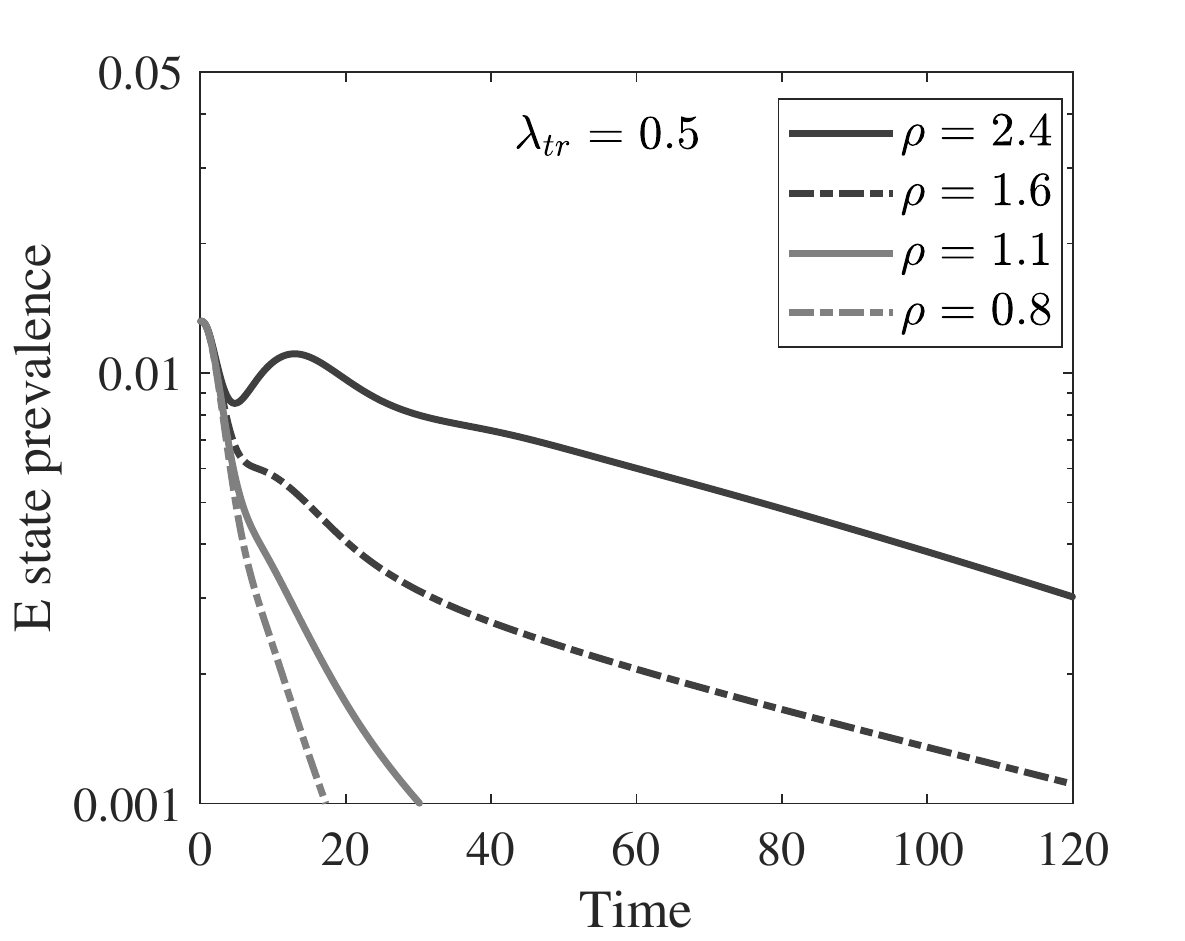}		
		\label{trcon2}}\\
		\subfloat[]{
		\includegraphics[width=0.24\textwidth]{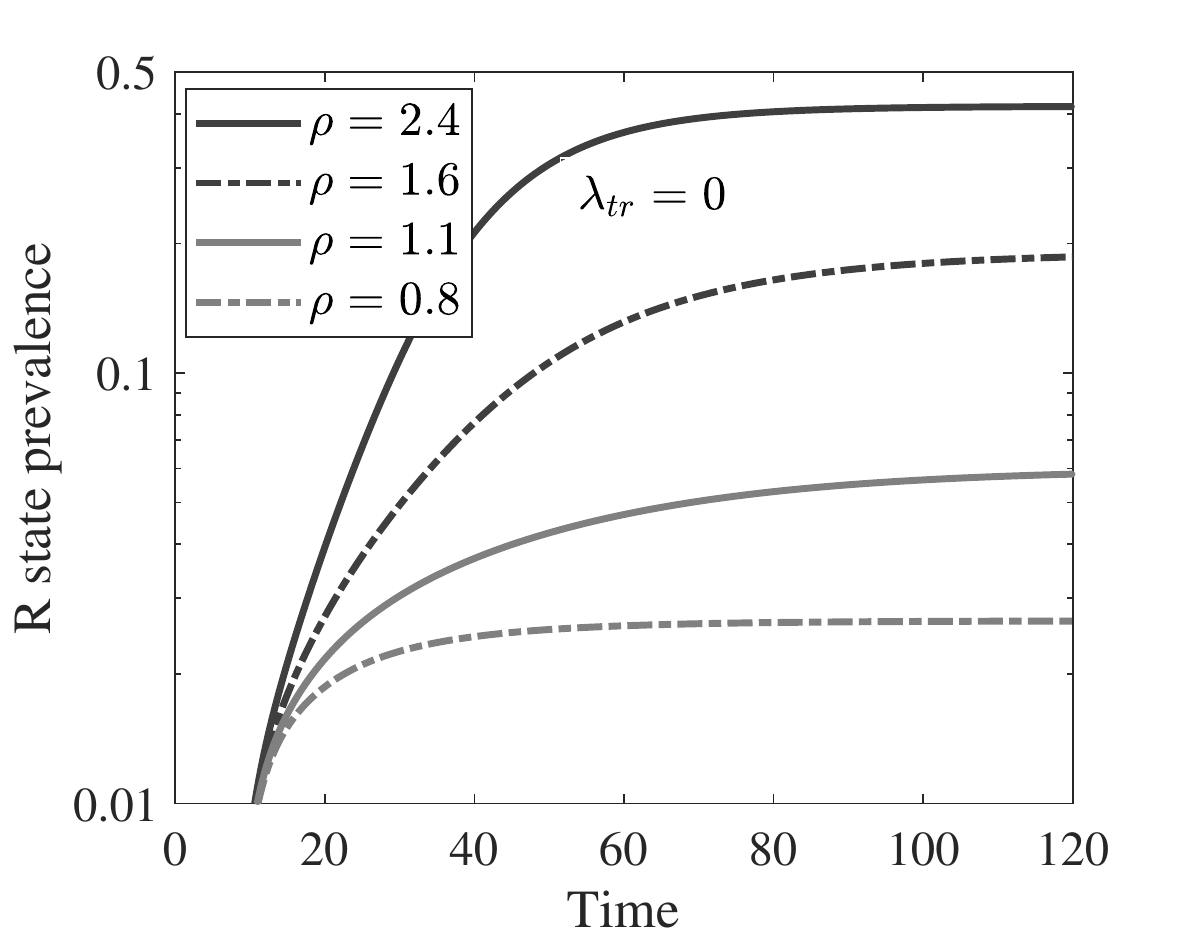}		
		\label{trcon3}} 
	\subfloat[]{
		\includegraphics[width=0.24\textwidth]{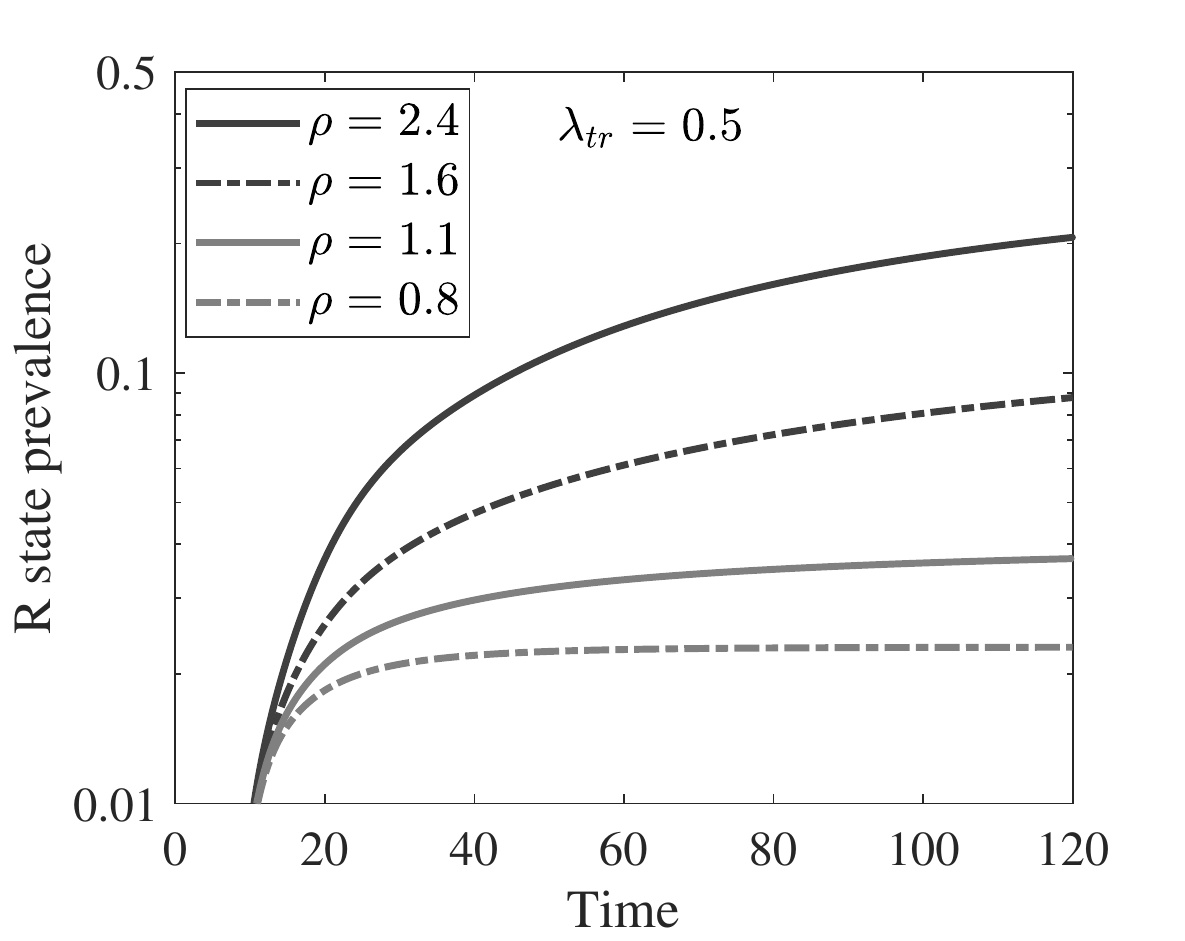}		
		\label{trcon4}}
	\caption{Prevalence of the E and R states calculated using the ODEs system \ref{dynamics} for an arbitrary network. For $\rho<1$, we expect the spreading to die out. } 
	\label{trcon}%
\end{figure}

To numerically investigate the threshold condition, we computed prevalence of the exposed state (E) through time for a spreading  unfolding on the largest component of the coauthorship network \cite{newman2006finding}. We calculated the prevalence of a state as the average of nodes' probabilities. Fig. \ref{trcon1} shows the prevalence of exposed state for different values of $\rho\left(-\mathbf{\Delta}\mathbf{\Sigma}^{-1} \right)$ when we assumed no contact tracing. Fig. \ref{trcon2} shows similar prevalence when there is contact tracing. In both figures, we can see the exposed state exponentially dies out when the spreading is below the threshold. To compere the size of epidemic for different values of $\rho\left(-\mathbf{\Delta}\mathbf{\Sigma}^{-1} \right)$, we have plotted the R state prevalence in Fig. \ref{trcon3} and \ref{trcon4}. 

\begin{figure}[t]
	\centering
	\subfloat[]{
		\includegraphics[width=0.24\textwidth]{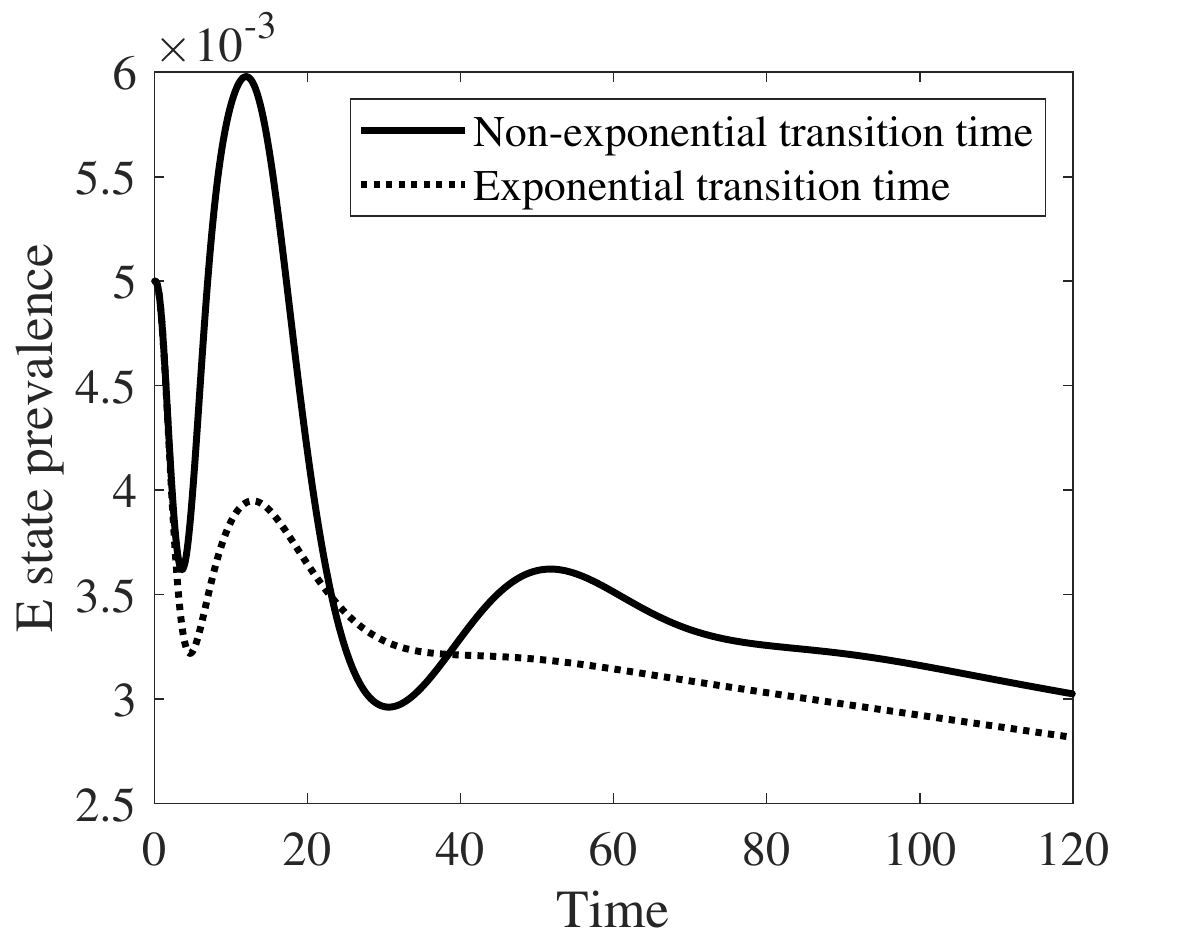}		
		\label{excomp1}} 
	\subfloat[]{
		\includegraphics[width=0.24\textwidth]{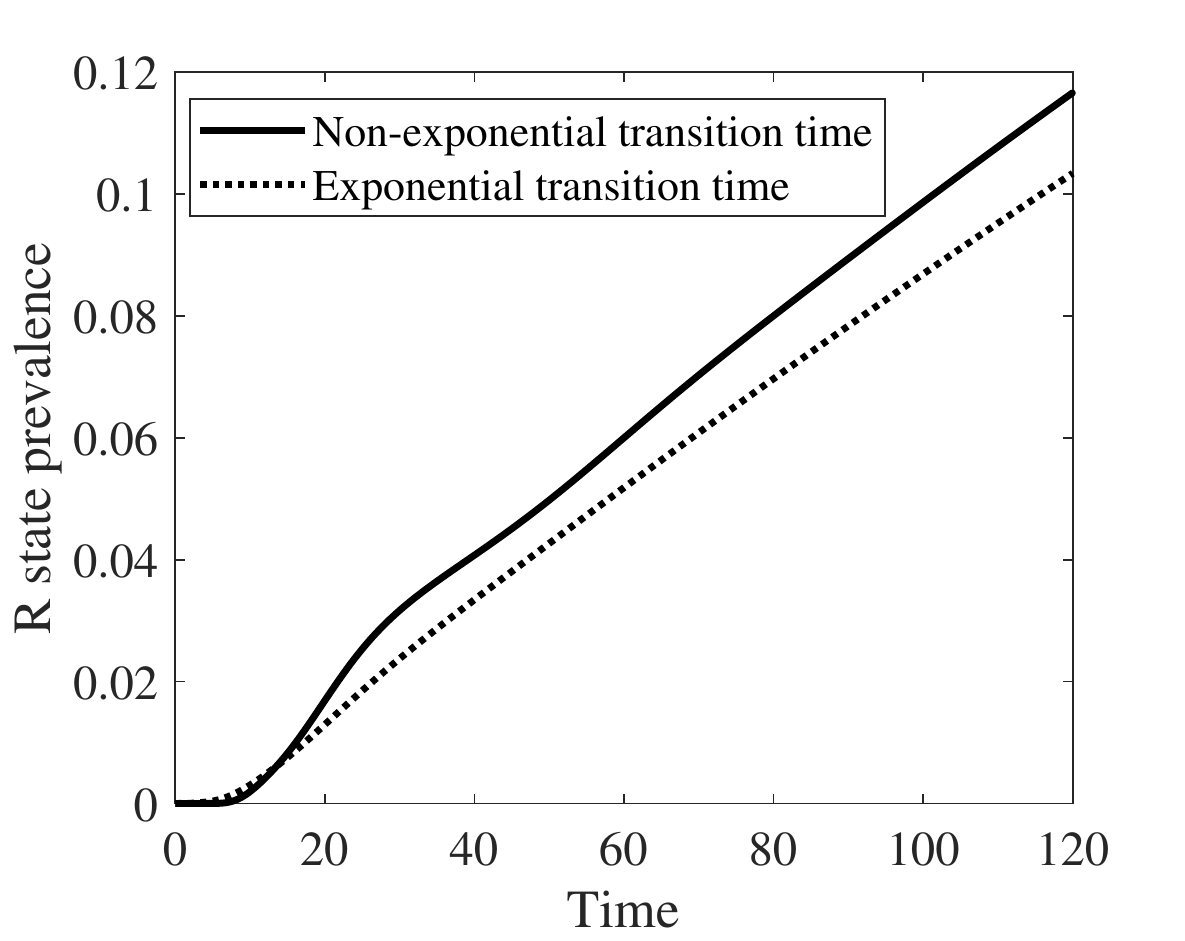}		
		\label{excomp2}}
	\caption{Effect of infectious period distributions on the E and R states' prevalences. For the curves tagged with "Non-exponential transition time", we used Erlang distributions with the same expected values as the the curves tagged with "Exponential transition time" but with different variances. } 
	\label{excomp}%
\end{figure}

The threshold condition we have derived, only depends on the network structure, infection transmission rates, probability of becoming a symptomatic case, and expected infectious periods for the symptomatic and asymptomatic states. This is clear from equation \ref{ttres}, if we notice that $k_{\Is}\lambda_{\Is}^{-1}$ and $k_{\Ia}\lambda_{\Ia}^{-1}$ are the expected values for the corresponding Erlang distributions. Although the variances for the infectious period distributions do not change the threshold condition, they affect the prevalence of the states as shown in Fig. \ref{excomp}. To generate this figure, we calculated the prevalence curves for two spreading processes unfolding on the network of the previous example. In the first process, we assumed the symptomatic and asymptomatic infectious periods are distributed exponentially with expected values of four and six, respectively. For the second process, we changed the exponential distributions to Erlang distributions with similar expected values but variance of 2.  

\section{Case study}\label{Case study}
 In this section the we apply the SAIDR spreading model to COVID transmission among Kansas State University (K-State) students in Manhattan, Kansas. Such a model can be used to evaluate the effectiveness of different strategies for reducing the spread of virus or to predict the size of epidemic. However, this requires information about the network structure and other model parameters. We performed a survey among individuals associated with the university and the survey results were used to build a random contact network. Assuming this contact network for the population, we use weekly positive COVID cases to estimate unknown model parameters such as the transmission rate $\beta_{Is}$. Later, we use the estimated parameters to study the effectiveness of contact tracing among the students.
 \subsection{Contact Network}\label{net} 
To build a weighted contact network we sent an online survey to all Kansas State University students and staff via the university webmail system in December 2020.  This survey asked about participation in social interactions during the 2020 fall semester (August 2020-December2020). During this semester, some of the classes were held in-person and some held online.  Students were present in Manhattan, Kansas, during this time and a mask mandate was in effect in all the facilities related to the university and in community public spaces. 
 
We asked questions about housing status, age and role in the university, as well as the number of close contacts such as roommates, family members, or coworkers. In another question we also asked for the number of people they regularly meet in close proximity for a total contact duration of less than four hours per week (such as friends). The responses to this question were used to build a second layer of contacts that can be traced but have lower transmission probability. To find the level of social interaction we asked the respondents to specify average number of visits per week to different public locations such as bars, restaurants, coffee shops and stores. We also asked about the frequency of their participation in social events such as religious or sports events. For each public location type, the respondents also indicated duration of the visits and number of people with whom they interact in close proximity.

After processing the data, based on their housing type and role in K-State, we divided the respondents into six groups. These groups are: (1) graduate students (2) factually and staff, (3) undergraduate students living in off campus apartment or houses, (4) undergraduate students living in fraternities and sorority houses (Greek houses), (5) and (6) undergraduate students using two available on-campus housing options. The rationale for this division is that the respondent age and housing type possibly leads to different levels of social interaction and number of close contacts. Using the survey data, we calculated the following parameters for each group $g$ 
\begin{itemize}
\item $\mathcal{V}_1^g$, average number of close contacts 
\item $\mathcal{V}_2^g$, average number of people met for total duration of less than four hours per week 
\end{itemize}
and for the three types of public spaces which are (1) bars, restaurants, and coffee shops, (2) stores and services, and (3) social events
\begin{itemize}
\item $\mathcal{P}_l^g$, proportion of individuals in group $g$ who visits public space type $l$, and  $\mathcal{H}_l^g$, average weekly hours they spent in these locations, and $\mathit{n}_l^g$, average number of people they encounter 
\end{itemize}
Values of these parameters are presented in the table \ref{table1}.
\begin{table}[h!]
\begin{center}
\begin{tabular}{ | c | c | c | c | c | c | c | } 
\hline
group \# & 1& 2 & 3& 4&5&6\\ 
\hline
$\mathcal{V}_1^g$& 3.5 & 3.5& 5.8&14.3&4.5&3.5\\ 
\hline
$\mathcal{V}_1^g$& 3.5 & 2.3& 4.9&10.7&5.1&4.2\\ 
\hline
location 1&  & & & & &\\ 
\hline
$\mathcal{P}_l^g$& 0.5 & 0.4 &0.6 & 0.73&0.53 &0.5\\ 
\hline
$\mathcal{H}_l^g$& 1.76 &0.7 & 3.25&4.46 & 2.25&1.5\\ 
\hline
$\mathit{n}_l^g$& 4 & 2.7&6 & 5.3& 5.3&3\\ 
\hline
location 2&  & & & & &\\ 
\hline
$\mathcal{P}_l^g$& 0.91 & 0.92&0.89 &0.81 & 0.75&0.91\\ 
\hline
$\mathcal{H}_l^g$& 2 & 1.7& 2.6& 2.1& 1.5&3\\ 
\hline
$\mathit{n}_l^g$& 4.5 & 4.2& 4.9& 4.3& 4.7&4.7\\ 
\hline
location 3&  & & & & &\\ 
\hline
$\mathcal{P}_l^g$& 0.3 & 0.26& 0.45& 0.72& 0.56&0.33\\ 
\hline
$\mathcal{H}_l^g$&0.85 & 0.7&1.72 & 4.5& 3&2.37\\ 
\hline
$\mathit{n}_l^g$& 6 & 5.5&8 & 14& 8.6&4.9\\ 
\hline
\end{tabular}
\end{center}
\caption{Value of the parameters introduced in section \ref{net}.}
\label{table1}
\end{table}

 Using these parameters, we built three layers of networks for the whole population comprising the groups mentioned before and an additional group which is the rest of the town population. Since we conducted the survey only among the individuals associated with university, we did not have the parameter values for the general public not associated with the university. Hence, we extended the parameters extracted for the group of faculty and staff to that additional group.
Population of different groups that we assumed in our calculations are given in the table below.
\begin{table}[h!]  
\begin{center}
\begin{tabular}{ | c | c | c | c | c | c | c | } 
\hline
group \# & 1& 2 & 3& 4&5&6\\ 
\hline
$N^{g}$ & 4000& 31000 & 8700& 2000&4400&900\\
\hline
\end{tabular}
\end{center}
\end{table}

For layer one, $\mathcal{L}_1$, which is the layer of close contacts, consider clusters of $\mathcal{V}_1^g+1$ individuals within  the main groups. We assume among $\mathcal{V}_1^g$ close contacts for each individual, a fraction $\mathcal{F}$ of the contacts happens within the cluster the individual belongs to, and the remaining contacts are randomly established using configuration model. While setting up the random links we assumed undergraduate students have only links with other undergraduate students and graduate students with other graduate students. 

For layer two, $\mathcal{L}_2$, we used configuration model where the node degree of an individual in group $g$ was set to $\mathcal{V}^g_2$.We assumed undergraduate students have only links with other undergraduate students and graduate students with other graduate students. To decrease the number of model parameters, we the express daily infection transmission rate in this layer in terms of transmission rate through the close contacts of layer $\mathcal{L}_1$, which we denote by $\beta$. If we assume the effective daily contact duration in $\mathcal{L}_1$ is only eight hours and compare that with weekly duration of links in $\mathcal{L}_2$, which is around four hours, we expect a daily transmission rate of  $4\beta/(7\times8)$ for the $\mathcal{L}_2$ links.

For the layer three, $\mathcal{L}_3$, which represents interaction through public spaces, we considered a complete graph over the whole population and weighted the link between any two nodes $i$ and $j$ by
\[
\sum_{l=1}^3 \mathcal{P}_l^{g(i)} \mathit{n}_l^{g(i)}\frac{\mathcal{P}_l^{g(j)} \mathit{n}_l^{g(j)}}{\sum_{k=1}^N\mathcal{P}_l^{g(k)} \mathit{n}_l^{g(k)}} \mathcal{H}_l^{g(i)} \mathcal{H}_l^{g(j)} \mathcal{C}_l
\]
In this expression, $l$ enumerates three different public spaces we mentioned before, $g(i)$ represents group assignment of node $i$ and $N$ is the total population. $\mathcal{C}_l$ is a coefficient that when is multiplied by $\mathcal{H}_l^{g(i)} \mathcal{H}_l^{g(j)} \beta$, gives an estimate of infection transmission rate between nodes $i$ and $j$ in public space $l$, assuming $\beta$ is transmission rate for a close contact link in the layer $\mathcal{L}_1$. Indeed, if $\mathcal{T}_{l}$ is the total hours per week that the public space is active  $\mathcal{H}_l^{g(i)} \mathcal{H}_l^{g(j)}/(7\mathcal{T}_{l})$ is daily expected hours that nodes $i$ and $j$ overlap in such a location. Moreover, if we assume the effective daily duration of close contact is eight hours, $\mathcal{H}_l^{g(i)} \mathcal{H}_l^{g(j)}\beta/(7\times8\mathcal{T}_{l})$  gives an estimate of daily infection transmission rate between nodes $i$ and $j$ in the public space $l$. Here, we use $\mathcal{T}_{1}=35$, $\mathcal{T}_{2}=70$, $\mathcal{T}_{1}=15$.
 
\subsection{Model Parameters}\label{pars}
For some of the parameters in the spreading model described in section \ref{analysis}, we use the values below in our calculations.

Following reference \cite{johansson2021sars}, we assume the proportion of infected individuals that never show symptoms is $p_{Ia}=0.30$, and their infectiousness is lower than those who develop symptoms such that $\beta_{\Ia}=0.75\beta_{\Is}$. 

\begin{figure}[t]
	\centering
	\subfloat[]{
		\includegraphics[width=0.24\textwidth]{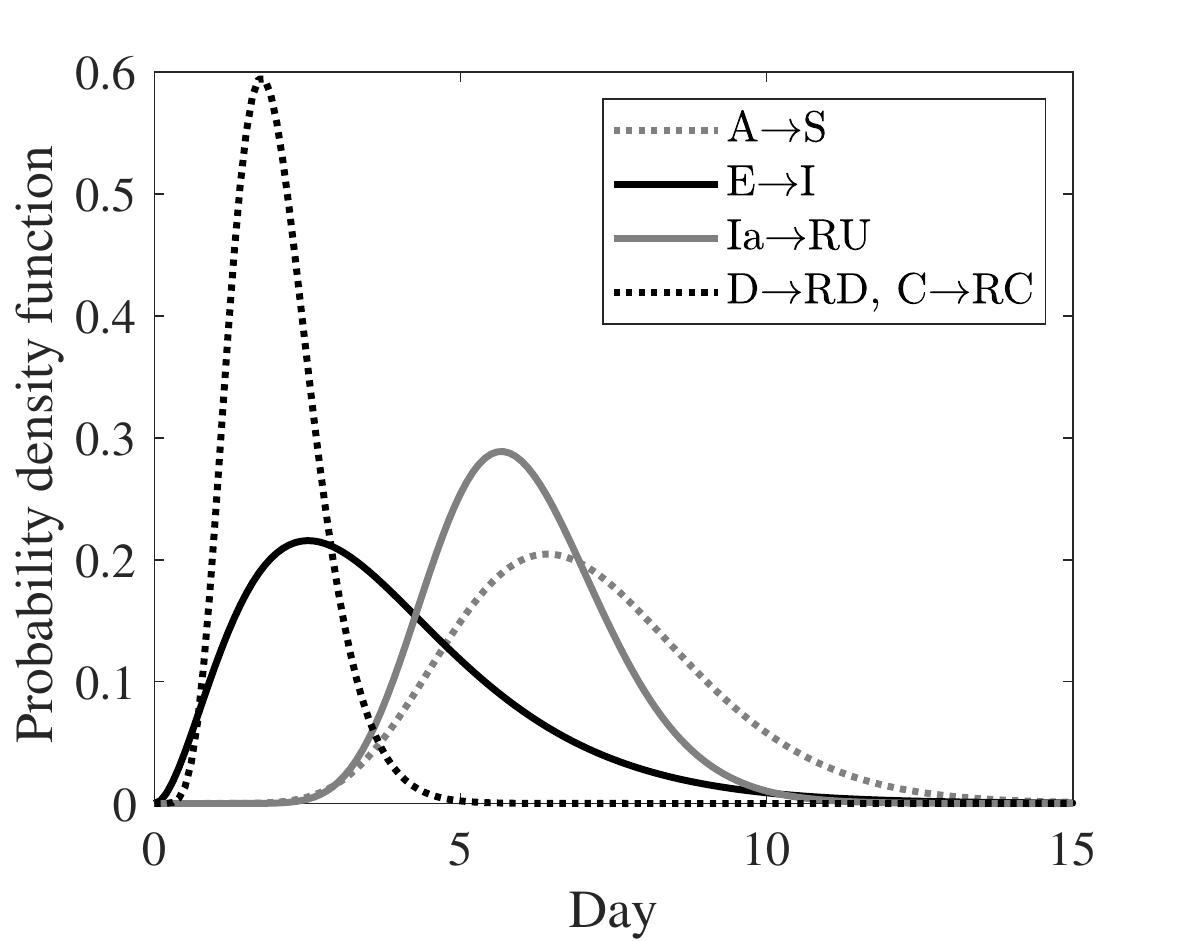}		
		\label{transpdf}} 
	\subfloat[]{
		\includegraphics[width=0.24\textwidth]{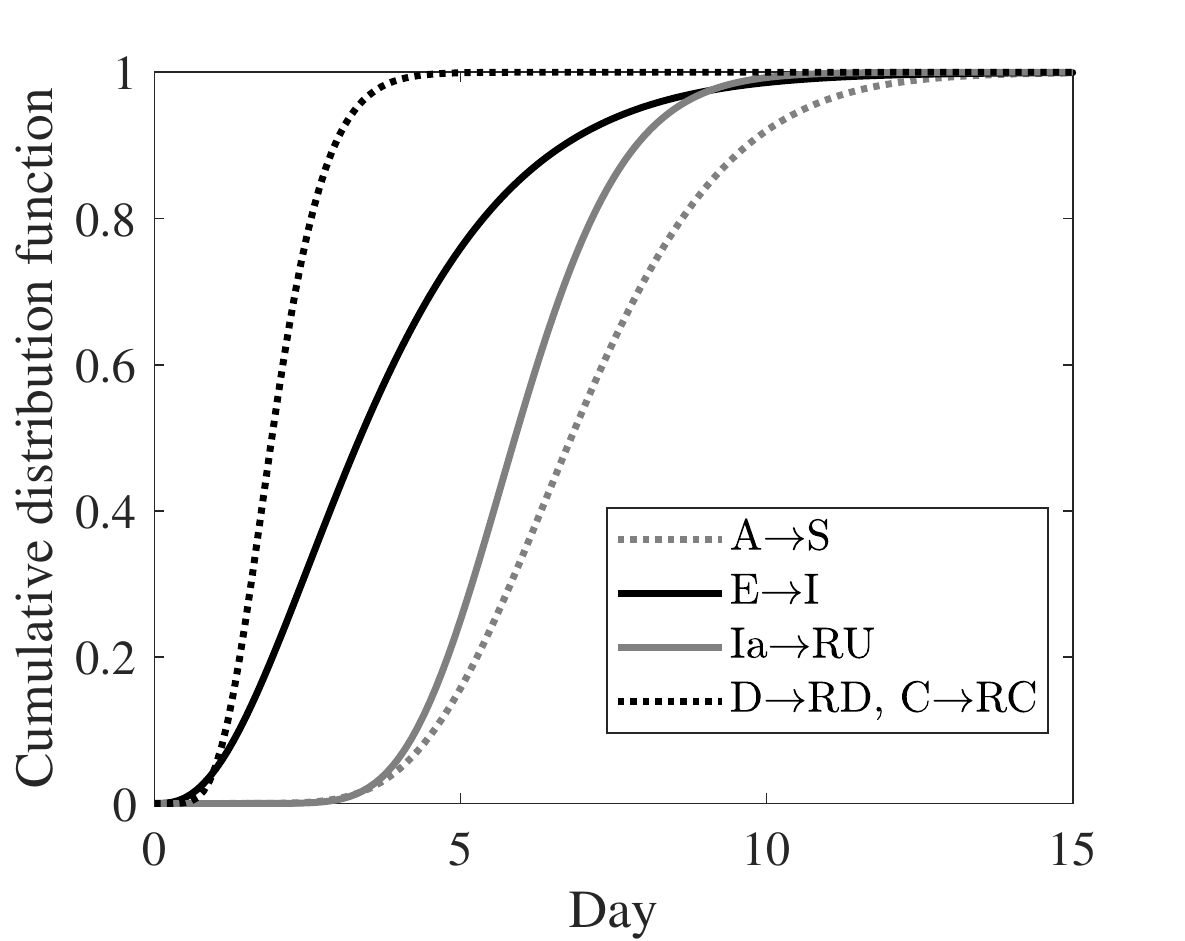}		
		\label{transcdf}}
	\caption{ Distribution of transition times used in our calculations.} 
	\label{transpcdf}%
\end{figure}

For transition from the E state to the I states, we use $\lambda_E^{-1}=1.255$ days, and $k_E=3$. These values lead to a distribution with the mean of $3.76$ days and the standard deviation of $2.17$ (Fig. \ref{transpcdf}). We chose this distribution using the data in reference \cite{ren2021evidence}, where authors report the distributions for the incubation period ( i.e., time from exposure to symptoms onset) and the serial interval which is time between the symptoms onsets of successive cases in a chain of transmission. In fact, the reported negative serial interval implies pre-symptomatic transmission of COVID-19 during the incubation period. To estimate the E state period in our model, we assumed that, in the incubation period, infected individuals go through several stages with the same expected time, and in one of the stages they start transmitting the infection. If this stage is before the onset of symptoms, we will observe negative serial intervals. Hence, we approximated the incubation period with an Erlang distribution with $k=4$ and $\lambda^{-1}= 1.255$ days, which has the same median and 95th percentile as the reported distribution in \cite{ren2021evidence}. Next, we simulated a chain of transmission assuming transmission starts at a specific stage of incubation period with a specific transmission rate, and we recorded the distribution of the serial interval. By exploring different values for the stage that the transmission starts, and also the transmission rate, we found that the simulated distribution of the serial interval is closest to the reported one in \cite{ren2021evidence}, if the transmission starts at the stage four. Therefore, we used an Erlang distribution with $k=3$ and $\lambda^{-1}= 1.255$ days for the period of E state in our model.

Considering the transition from the $\Ia$ state to the RU state, we assumed an Erlang distribution of mean six days and variance of two days. Fig. \ref{transpcdf} shows the corresponding density function. This value of variance implies that there is almost no transmission after day nine \cite{cevik2020sars} and there is transmission in the first week when the viral load is high \cite{cevik2020sars}.

Here we use similar distributions for the random times of the of the transitions C$\rightarrow$ RC and D$\rightarrow$ RD. If we assume $k_C=k_D=8$ and $\lambda_C^{-1}=\lambda_D^{-1}=0.25$ days, and $\lambda_{tr}=1.5\ \text{day}^{-1}$ , then the probability that a node in $D$ or $C$ state induces a contact tracing transition in its neighbor is $0.92$.
We calculate the contact tracing success probability as
\begin{equation}
\begin{split}
&\int_0^\infty f(t|k_{D},\lambda_{D})(1-e^{-\lambda_{tr}t})dt\\
&=\frac{\lambda_{tr}}{\lambda_{tr}+\lambda_{D}}\sum_{i=0}^{k_D-1}\left(\frac{\lambda_{D}}{\lambda_{tr}+\lambda_{D}}\right)^i.
\end{split}
\end{equation}  
    We can adjust this probability by changing $\lambda_{tr}$ or  $k_C,\ k_D,\ \lambda_C,\ \lambda_D$. For instance, if we change $\lambda_{tr}$ to only $0.18\ \text{day}^{-1}$, the probability becomes $0.3$. In our calculations, we assume the rate of success for contact tracing through the layer of close contacts, $\mathcal{L}_{1}$ is high, and through the layer $\mathcal{L}_{2}$ is lower, therefore, we chose $\lambda_{tr}=1.5 \text{ and}  \ 0.18\ \text{day}^{-1}$ for the layers $\mathcal{L}_{1}$ and $\mathcal{L}_{2}$, respectively. We need to note that the duration of staying in the C or D state cannot be long, otherwise these states may induce repetitive contact tracing in the model. 
 
The other transition that we specify its random time distribution is A$\rightarrow$ S. Here, we set the $k_A=12$ and $\lambda_A^{-1}=0.5833$ days. These values lead to a distribution that is concentrated between days 4 and 11 with a mean of 7 days (Fig. \ref{transpcdf}).  

 \subsection{Approximate Model}\label{apm}
 Practically, we can solve the network spreading model ODEs of the system \ref{dynamics}, even if the population is large. This system of equations is applicable for any network with an arbitrary structure. However, the three-layer network we described in section \ref{net}, to some extent, is homogeneous. Specifically, the nodes that belong to a same group have similar type and number of links in the layers $\mathcal{L}_1$ and $\mathcal{L}_2$, and their community contacts through the layer $\mathcal{L}_3$  are identical. Hence, we expect that the probability vectors of the nodes with a similar group assignment will be similar, and we may use one probability vector for all the nodes in a group. 
Within this approximation, the dimension of the dynamical system that describes the spreading process is smaller, and each group of nodes is represented by only one node. The ODEs for this system are similar to those in the equation \ref{dynamics}, except that the superscript $i$ for the variables now enumerate the groups and run from $1$ to $6$. To have a closed system of ODEs, we also need to approximate the infection and tracing rates, $r_I^i$ and $r_T^i$ in term of the probability vectors of the groups. The infection rate of a node in group $i$ can be written as
\begin{equation}
r^{i}_{I}= \sum_{j=1}^6\Omega_{1}^{i,j} \left(\beta_{\Is}\sum_{k} \Is ^{j}_{k}+\beta_{\Ia}\sum_{k}\Ia^{j}_{k} \right),
\end{equation}
where $\Omega_{1}^{i,j}$ determines the contribution of group $j$ in the infection rate of a node in group $i$ and $\beta_{\Is}$, $\beta_{\Ia}$ are infection transmission rates through close contacts of the layer $\mathcal{L}_1$. Considering definition of the network parameters in section \ref{net}, $\Omega_{1}^{i,j}$  is approximated by 
\begin{equation}\label{omega1}
\begin{split}
\Omega_1^{i,j}&=\mathcal{V}_1^i\left(1-\mathcal{F}\right)\frac{\mathcal{V}_1^j\left(1-\mathcal{F}\right)\omega^{i,j}N^j}{\sum_k \mathcal{V}_1^k\left(1-\mathcal{F}\right)\omega^{i,k}N^k}+\delta^{i,j}\mathcal{F}\mathcal{V}_1^i\\
&+\zeta_1\mathcal{V}_2^i\frac{\mathcal{V}_2^j\omega^{i,j}N^j}{\sum_k \mathcal{V}_2^k\omega^{i,k}N^k}\\
&+\sum_{l=1}^3 \mathcal{P}_l^{i} \mathit{n}_l^{i}\frac{\mathcal{P}_l^{j} \mathit{n}_l^{j}N^j}{\sum_{k=1}\mathcal{P}_l^{k} \mathit{n}_l^{k}N^k} \mathcal{H}_l^{i} \mathcal{H}_l^{j} \mathcal{C}_l
\end{split}
\end{equation}
where $\delta^{i,j}$ is the Kronecker delta function, $N^i$ is the population of group $i$ and $\zeta_1=4/(7\times 8)$ is the ratio of infection transmission rate for the contacts in the layer $\mathcal{L}_2$ to that of the layer $\mathcal{L}_1$. Furthermore, $\omega^{i,j}=1$, if existence of links between the nodes in groups $i$ and $j$ is allowed in the network layers $\mathcal{L}_1$ and $\mathcal{L}_2$, otherwise $\omega^{i,j}=0$. The first, second and third lines in the equation above give the contribution of the network layers $\mathcal{L}_1$, $\mathcal{L}_2$ and $\mathcal{L}_3$ in the infection rate, respectively. Also, the contact tracing rate for the nodes in group $i$, represented by $r_T^i$, can be approximated as   
 \begin{equation}
r^{i}_{T}= \sum_{j}\Omega_{2}^{i,j} \lambda_{tr} \left(\sum_{k}C^{j}_{k}+\sum_{k}D^{j}_{k} \right),
\end{equation}
where $\lambda_{tr}=1.5\ \text{day}^{-1}$ is the tracing rate through the close contacts in the layer $\mathcal{L}_1$ and
\begin{equation}
\begin{split}
\Omega_2^{i,j}&=\mathcal{V}_1^i\left(1-\mathcal{F}\right)\frac{\mathcal{V}_1^j\left(1-\mathcal{F}\right)\omega^{i,j}N^j}{\sum_k \mathcal{V}_1^k\left(1-\mathcal{F}\right)\omega^{i,k}N^k}+\delta^{i,j}\mathcal{F}\mathcal{V}_1^i\\
&+\zeta_2\mathcal{V}_2^i\frac{\mathcal{V}_2^j\omega^{i,j}N^j}{\sum_k \mathcal{V}_2^k\omega^{i,k}N^k}
\end{split}
\end{equation}
In the equation above, $\zeta_2$ is the ratio of tracing rate through the layer $\mathcal{L}_2$ to $\lambda_{tr}$, which we set at $0.18/1.5$.

 \begin{figure}[t]
	\centering
	\subfloat[]{
		\includegraphics[width=0.24\textwidth]{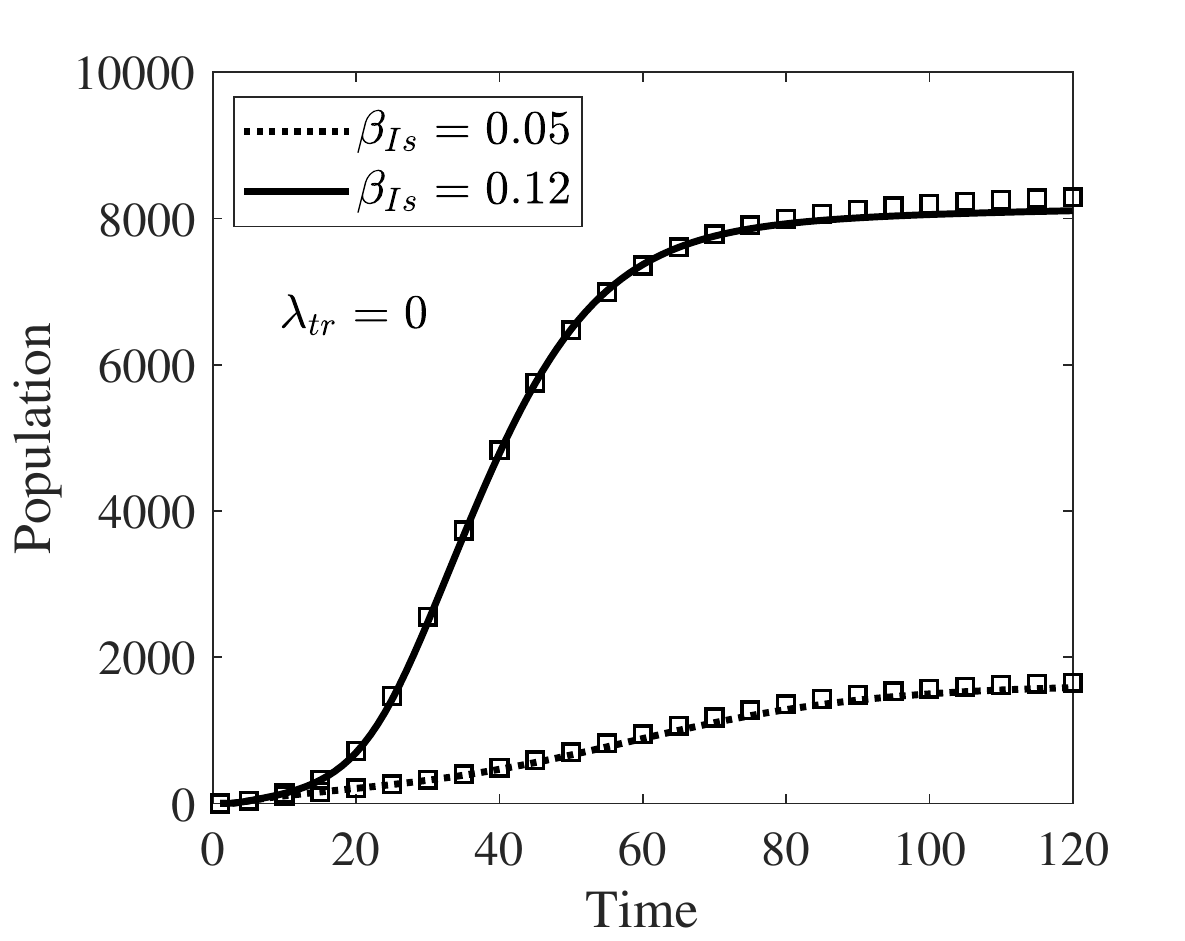}		
		\label{apne1}} 
	\subfloat[]{
		\includegraphics[width=0.24\textwidth]{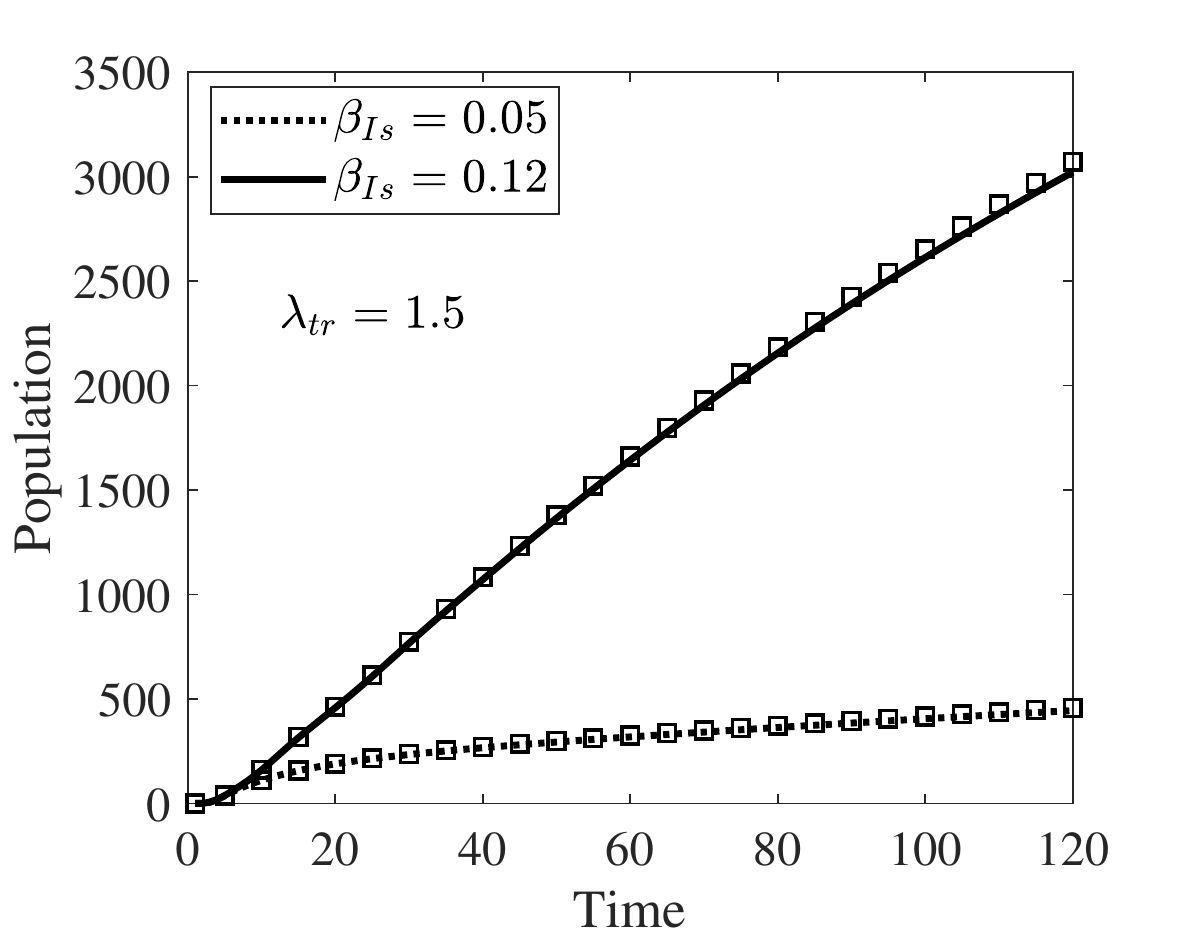}		
		\label{apne2}}
		 
	\caption{Comparing the results of the approximate model of section \ref{apm} (shown by solid lines) with the network model ODEs system \ref{dynamics} (shown by square markers).}
	\label{apne}%
\end{figure}

To compare the approximated spreading model with the high dimension network model, we calculated the total population of students in the D, C, RD, RC states using  both models assuming similar initial condition and the contact network parameter $\mathcal{F}=0.3$. Fig. \ref{apne} shows this population for different values of $\beta_{Is}$ and $\lambda_{tr}$. In this figure, the points shown by square markers were calculated using the network spreading model equations and the line plots are the result of the approximate spreading model. We can see, for the type of contact networks we defined in this section, the approximate model generates epidemic curves comparable to those obtained by the network spreading model.

\subsection{Estimation of Effectiveness of Contact Tracing}
In this section we use the reported positive COVID cases to estimate the effectiveness of contact tracing among the students during the 2020 fall semester. We assume the infection spreading follows our spreading model, and the contact network and the parameters' values are those discussed before in sections \ref{net}, \ref{pars}. Since we do not know the value of infection transmission rate $\beta_{\Is}$ and the distribution of infectious period for the symptomatic state, Is, we use a Markov chain Monte Carlo (MCMC) scheme to estimate these unknown parameters and eventually obtain an estimation for the effectiveness of the contact tracing. Although there are some published results regarding these parameters, we believe they cannot be extended to all populations. For instance, the distribution of the symptomatic infectious period, by which we mean the period an infectious symptomatic is spreading infection before removal from the population, depends on the level of population vigilance and testing. 

\begin{figure}[t]
	\centering
			\subfloat[]{
		\includegraphics[width=0.24\textwidth]{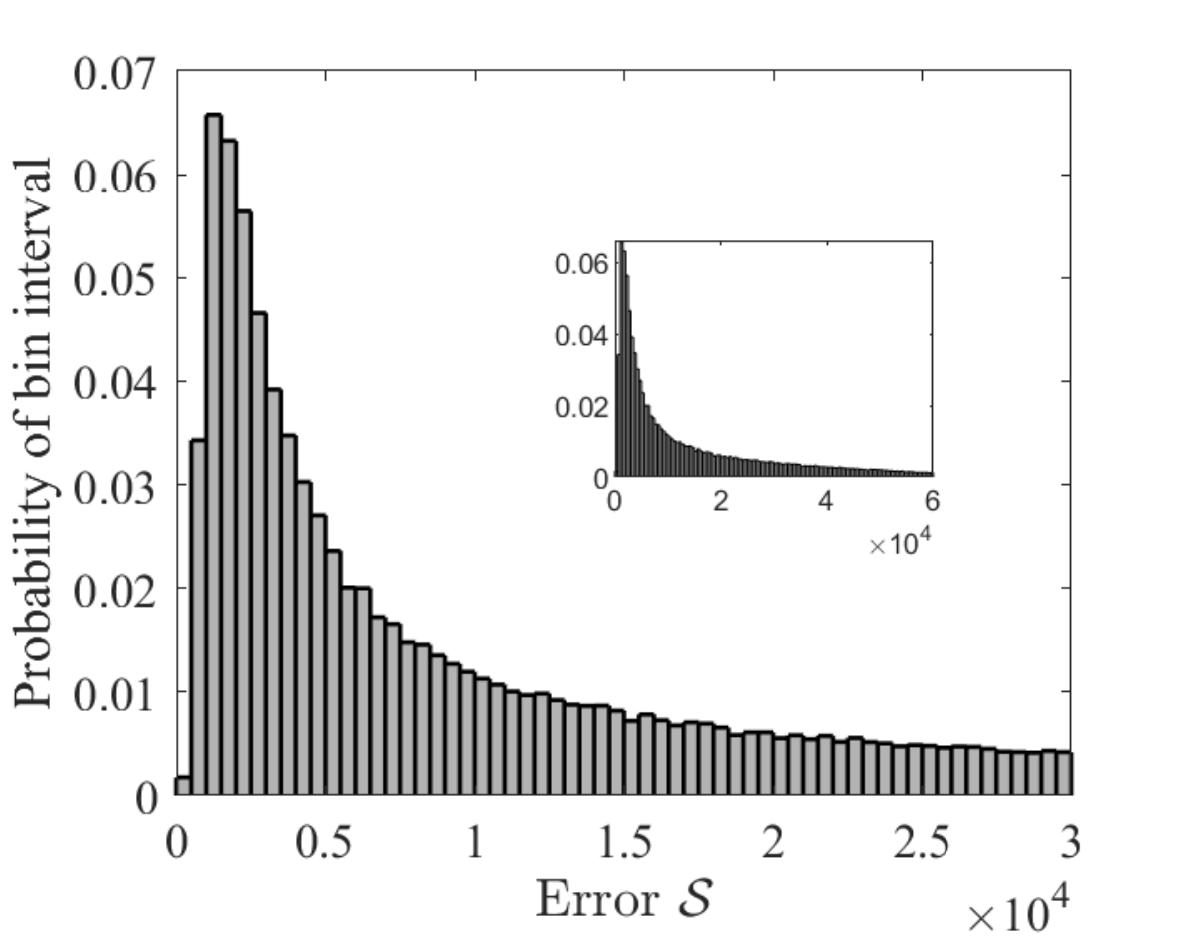}		
		\label{sam1}} 
	\subfloat[]{
		\includegraphics[width=0.24\textwidth]{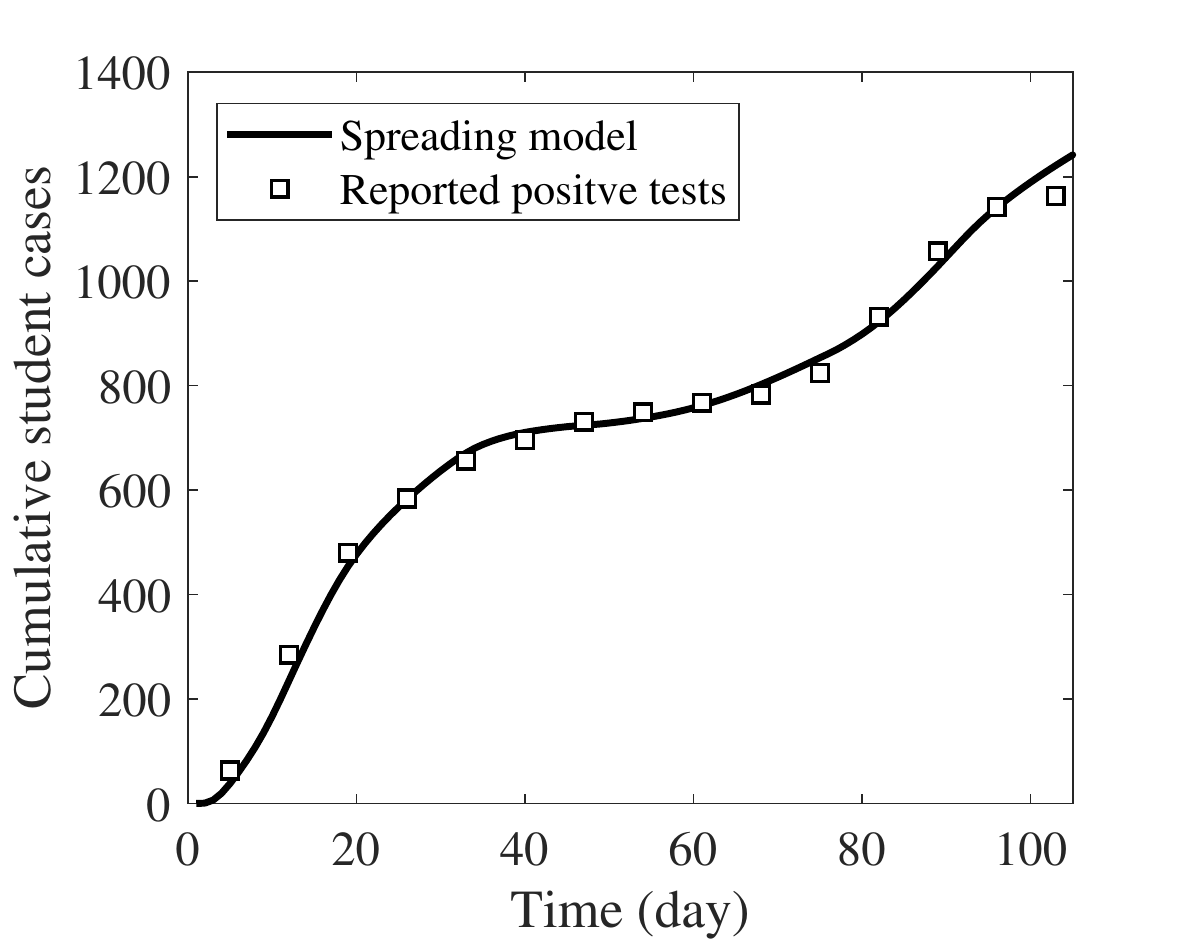}		
		\label{sam2}}\\
	\subfloat[]{
		\includegraphics[width=0.24\textwidth]{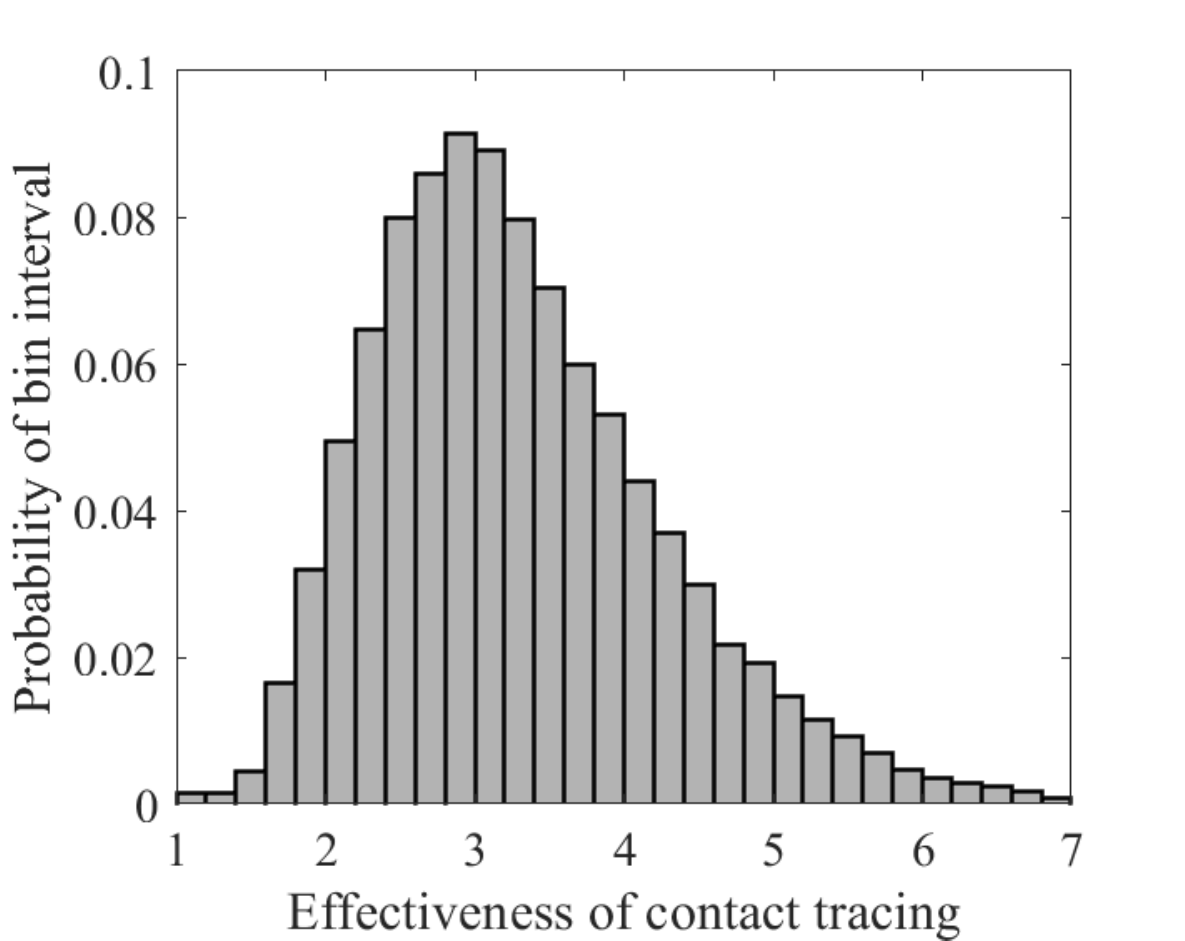}		
		\label{sam3}} 
	\subfloat[]{
		\includegraphics[width=0.24\textwidth]{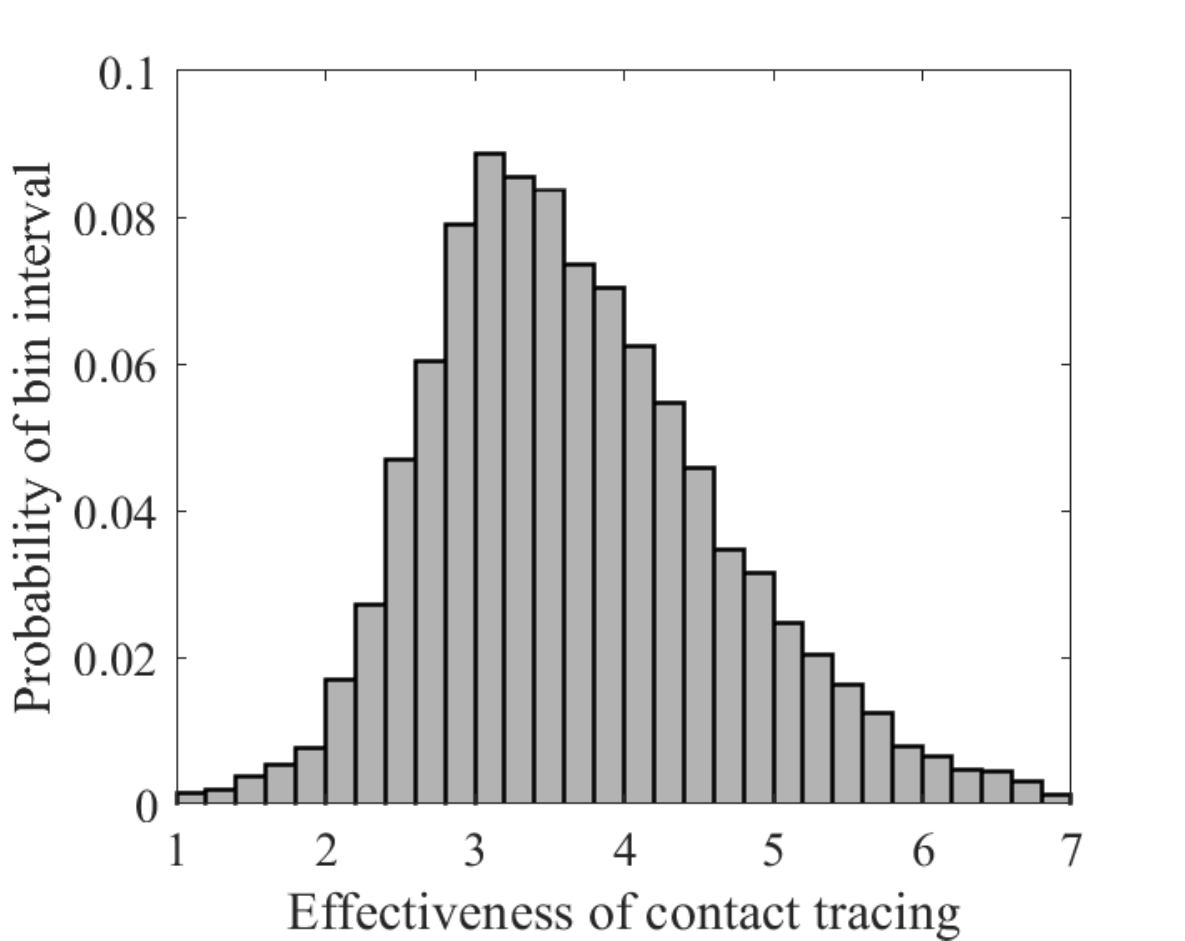}		
		\label{sam4}}

	\caption{Plot (a) Shows histogram of the MCMC samples' error. Plot (b) shows the epidemic curve calculated using the MCMC sample that has the minimum error. Panel (c) shows the histogram of contact tracing effectiveness for the MCMC samples and panel (d) shows a similar histogram when the samples with large error are ignored.} 
	\label{sam}%
\end{figure}

Here we use the Metropolis–Hastings algorithm, which is a MCMC method, to estimate the spreading model parameters. In general, this algorithm generates samples of a random variable for which we can calculate the probability ratio between the samples. Later, these samples can be used for calculating numerical approximations of functions of the random variable.
To detail the algorithm, consider a k-dimensional random vector ${\displaystyle \mathbf {X} =(X_{1},\ldots ,X_{k})}$, with a probability distribution proportional to $f(x_{1},\ldots ,x_{k})$. The algorithm generates a sequence of $M$ random samples $\{\mathbf{x}^t=(x_{1}^t,\ldots ,x_{k}^t)\}_{t=1}^M$ following the steps below
\begin{enumerate}
\item Choose an initial sample $\mathbf{x}^1$, and set $t=1$
\item Generate a proposal sample $\mathbf{x}=(x_{1},\ldots ,x_{k})$, such that each $x_{i}$ is normally distributed with a mean of $x_{i}^t$ and a predefined standard deviation, i.e., $x_{i}\sim \mathcal{N}(x_{i}^t,\sigma^2_i)$
\item Calculate acceptance probability, $\rho=\min\left(1, \frac{f(\mathbf{x})}{f(\mathbf{x}^t)}\right)$
\item Generate a uniformly distributed random number in the interval $[0,1]$. 
\item If $r\leq\rho$ accept $\mathbf{x}$ as the next sample, $\mathbf{x}^{t+1}=\mathbf{x}$, otherwise set $\mathbf{x}^{t+1}=\mathbf{x}$.
\item Set $t=t+1$, and go back to step 2.
\end{enumerate}
In order to adopt this algorithm in our estimation problem, we need to define the function $f(\mathbf{x})$ for our problem. Let $\{o_s\}$ and $\{y_s\}$ denote an observed spreading data series and the corresponding  outputs of spreading model. For a vector of spreading model parameters, $\mathbf{x}$, we assume $f(\mathbf{x})=\mathcal{S}(\mathbf{x})^{-2}$, where 
$\mathcal{S}(\mathbf{x})$ is the sum of absolute deviation of the spreading model outputs, $\mathcal{S}=\sum_s\mid o_s-y_s\mid$. Using this function, the fitting algorithm generates a sequence of samples from the parameter space such that the density of samples in  a region with the error $\mathcal{S}$ is four times the 
density in a region with the error $2\mathcal{S}$. In the definition of $f(\mathbf{x})$, we have used the sequences $\{o_s\}$ and $\{y_s\}$, which we need to specify. Since we had access to the weekly new positive COVID cases among the students, we use the weekly cumulative cases from the start of the fall semester as  the sequence of observed spreading data $\{o_s\}$. The corresponding sequence from the model, $\{y_s\}$, is the weekly population of students in the states D, DR, C, CR. Regarding the spreading model, we use the approximate model in section \ref{apm} because it is computationally fast and its outputs are similar to those calculated by the network model ODEs \ref{dynamics}. Concerning the open parameters in spreading model, we assumed the parameter vector $\mathbf{x}$ consists of the following components: 
\begin{itemize}
\item Mean and variance of the infectious period for the symptomatic state. Assuming these parameters, we find an Erlang distribution with same mean and a variance closest to the parameter value, and we use this distribution in the model.
\item Transmission rate $\beta_{\Is}$, which we allow to change every 15 days. Effectively, through the course of spreading, average $\beta_{\Is}$ in every 15 days is represented by one component of $\mathbf{x}$.
\item Timepoint of the first observation in the sequence $\{o_s\}$ with respect to the initial day of the model calculation. We solve the ODEs for the model with the initial E state probability equal to $0.005$.
\item Parameters $\mathcal{F}$ which is defined in section \ref{net}, and relates to the structure of the close contact network layer.
\end{itemize}
Following the procedure we explained in this section, we generated sequences with 1 million samples of the open parameters. Fig. \ref{sam1} shows a histogram of the samples' absolute error $\mathcal{S}$. The output of the spreading model for a sample with the minimum error of 330 is shown in 
Fig. \ref{sam2} where we have plotted the population of students in the D, C, RD and RC states and compared it with the reported students' positive cases.  

To estimate the effectiveness of contact tracing, we set the contact tracing rate in the model to zero, $\lambda_{tr}=0$, and for each sample of the open parameters we recalculated the total population of infected students at the day of last observation. Next, we calculated the ratio of this population to the same population in the original model where $\lambda_{tr}=1.5\ \text{day}^{-1}$, and we use this ratio to quantify effectiveness of contact tracing. Histogram of this effectiveness is shown in Fig. \ref{sam3}. When we examine the distribution of samples error in Fig. \ref{sam1}, we notice the distribution has a long tail. This is because, in the sampling algorithm, we chose $f(\mathbf{x})=\mathcal{S}(\mathbf{x})^{-2}$. In fact, we could instead use an exponential function that would not generate a long tail distribution, but the resulting Markov chain could possibly stay in a local minima for a long time. To eliminate the effect of samples with large errors, in our estimation of parameters, we only use samples with error $\mathcal{S}$ smaller than $5000$. This effectively truncates the original distribution. Fig. \ref{sam4} shows the distribution of contact tracing effectiveness when we only use samples with error less than $5000$. The mean for this distribution is 3.35 which is slightly different from that of the distribution in Fig. \ref{sam3} which is 2.94.

\begin{figure}[t]
	\centering
			\subfloat[]{
		\includegraphics[width=0.24\textwidth]{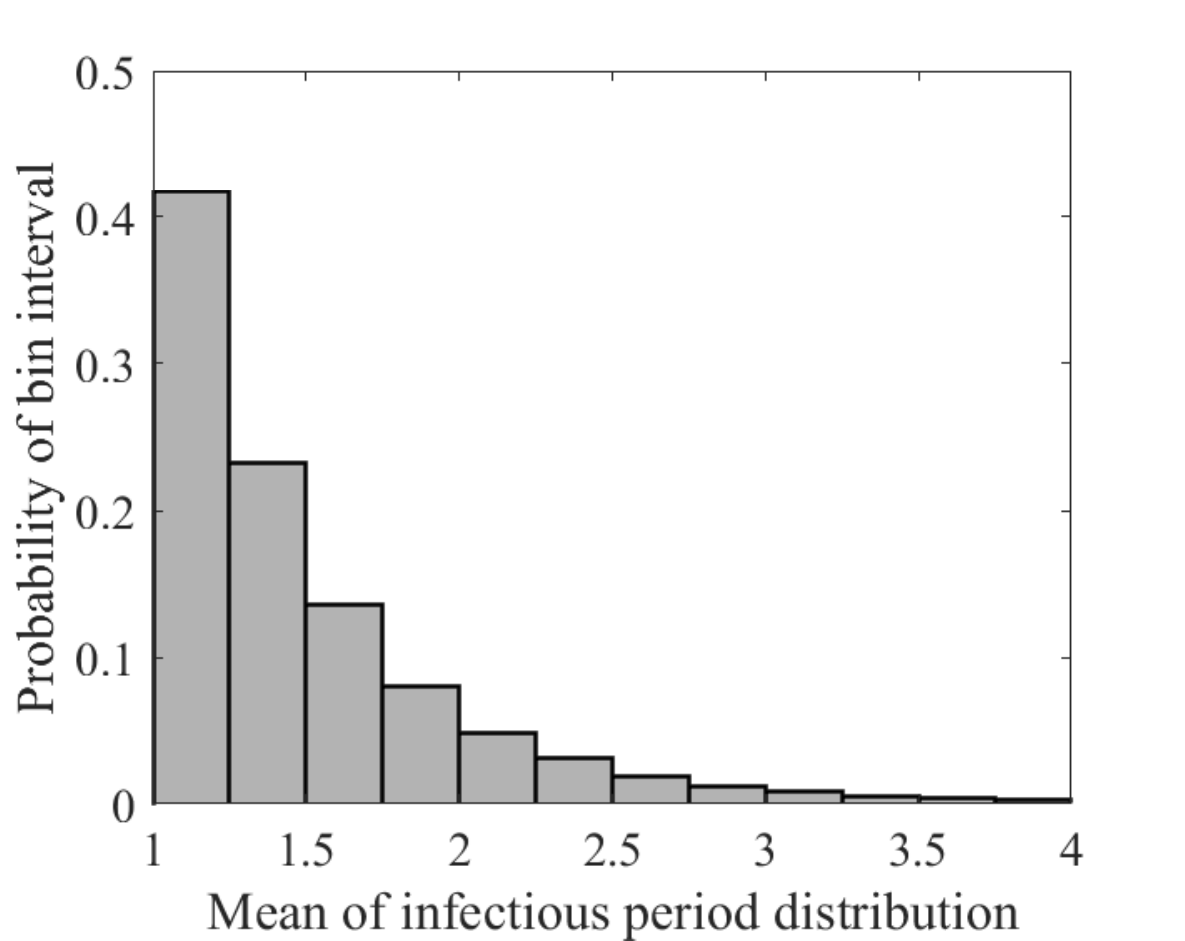}		
		\label{es1}} 
	\subfloat[]{
		\includegraphics[width=0.24\textwidth]{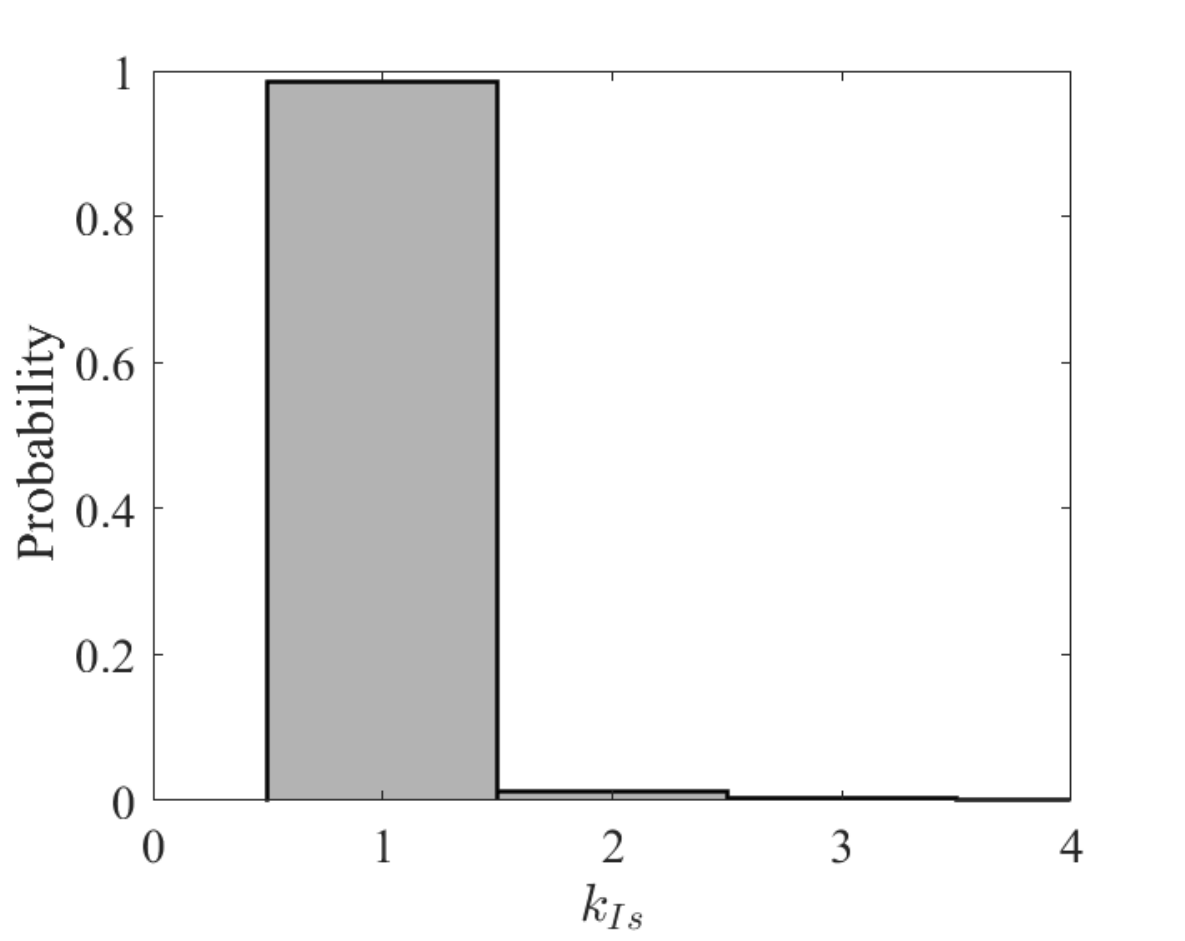}		
		\label{es2}}\\
	\subfloat[]{
		\includegraphics[width=0.24\textwidth]{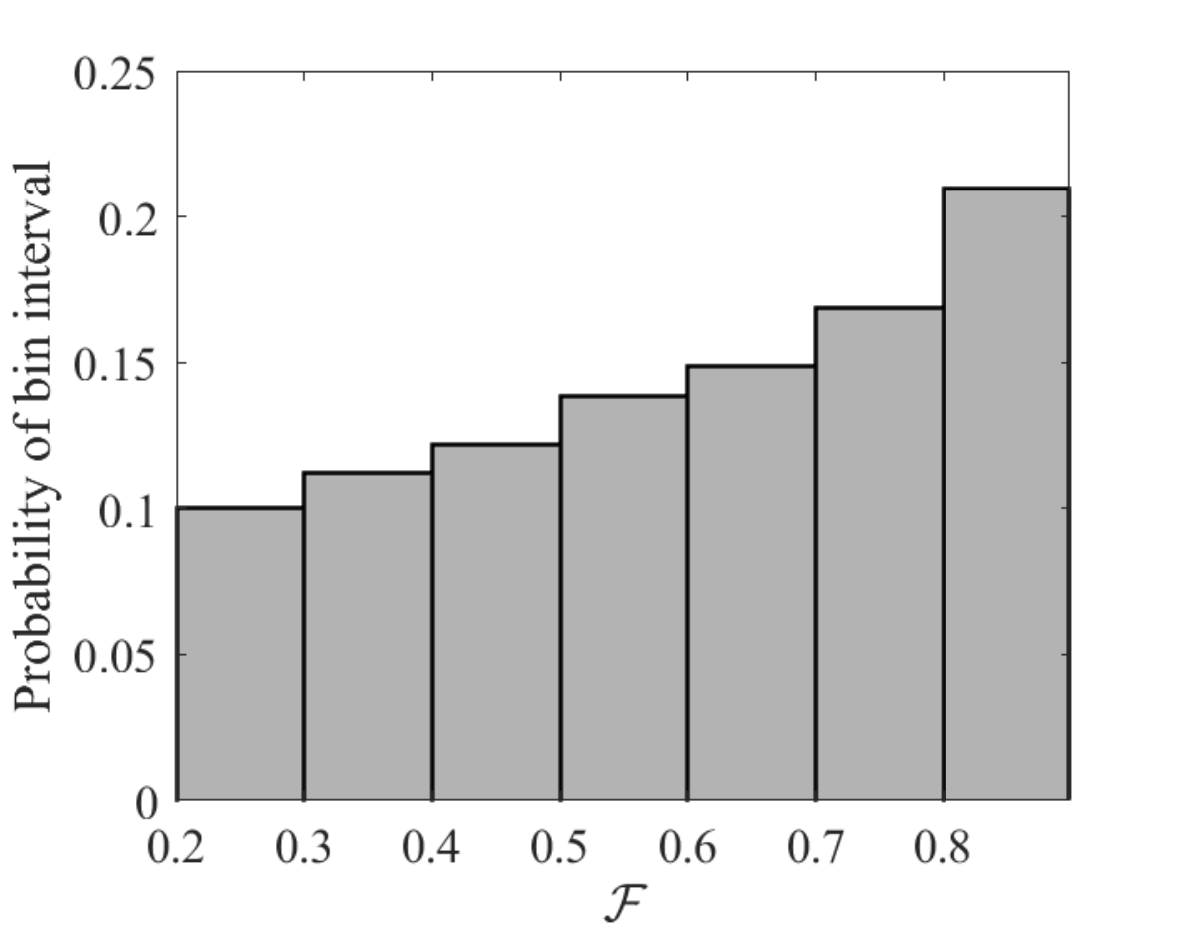}		
		\label{es3}} 
\subfloat[]{
		\includegraphics[width=0.24\textwidth]{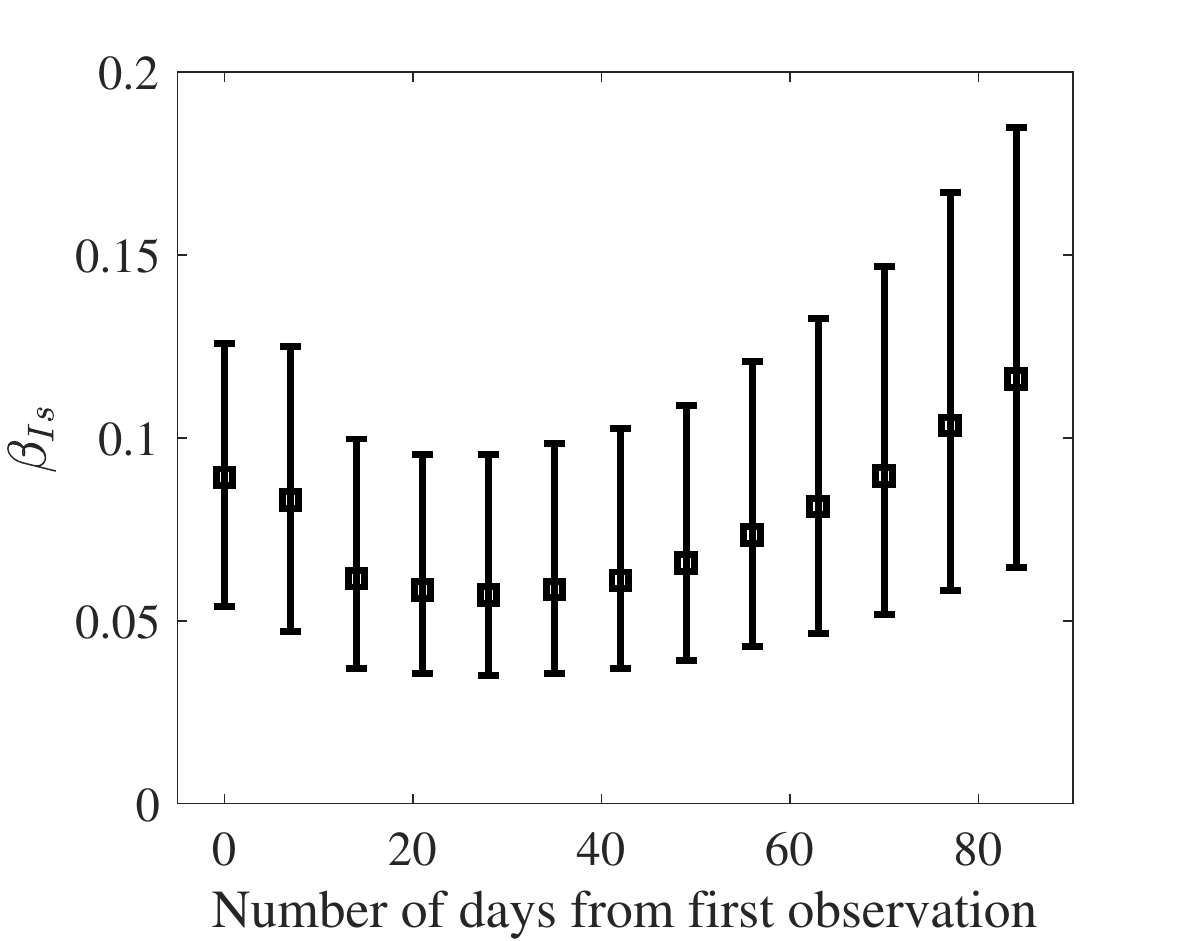}		
		\label{es4}} 

	\caption{Histograms of the model's open parameters extracted from the MCMC samples.} 
	\label{es}%
\end{figure}

Moreover, from each sample we can extracted  $k_{\Is}$ for the corresponding Erlang distribution of the infectious period. The histogram of  
$k_{\Is}$  is plotted in the Fig. \ref{es2}. This shows the most probable distribution for the duration of the Is state is exponential. We have also shown the distribution of the infectious period mean for the Is state in Fig. \ref{es1}. The mean for this distribution is $1.5$ days which is significantly shorter than the infectious period for the Ia state. The reason for this might be the vigilance of population and availability of testing. Fig. \ref{es3} shows the distribution for the network parameter $\mathcal{F}$, which has mean of $0.6$. We have also shown the Interquartile range and the estimated median for $\beta_{\Is}$ in Fig. \ref{es4}. In this plot we can see a range of values for $\beta_{\Is}$, but if we assume $\beta_{\Is}=0.1$, the factor $\left(\beta_{\Is} p_{\Is}k_{\Is}\lambda_{\Is}^{-1}+ \beta_{\Ia} p_{\Ia}k_{\Ia}\lambda_{\Ia}^{-1}\right)$ in the threshold equation \ref{ttres} becomes $0.24$. Considering the fact that the largest eigenvalue of a network is larger than its minimum node degree, we can deduce that if the individuals in a population have more than four close contacts the threshold condition for the epidemic die-out will not satisfy. This is especially significant because in the calculation we have assumed the mean of infectious period for the Is state is only 1.5 days which is very small.   
 
\section{Conclusions}
In this work, we have developed and investigated a novel non-Markovian network-based model to assess the effectiveness of contact tracing and shed light on conditions to contain a COVID-19 outbreak. In particular, using equation \ref{threshold} for the epidemic threshold, it is clear the quantitatively relevant impact of the asymptomatic infected individuals in the spreading process. Furthermore, when considering homogeneous contact patterns, even limiting the number of contacts to a very small number (i.e.,4-5) per individual will not be enough to contain the epidemic. We have evaluated contact tracing with a two-layer network, a regular-contact layer and a casual-contact layer. Extensive simulations show that contact tracing can be very effective by reducing the size of the epidemic more than three-fold. This result takes into account that 90\% of the regular closed contacts are traced and only 30\% of the casual contacts are traced. In our model, the duration of the symptomatic and asymptomatic infection periods are evaluated at 1.5 days mean for symptomatic individuals due to prompt isolation, while at 6 days for asymptomatic not-detected ones. Since the threshold equation is derived using a mean-field first-order closure approximation, future work will explore higher order closure approximations evaluating the trade-off between improved accuracy and increased costs. Overall, we have proposed a non-Markovian model for evaluating SARS-CoV-2 transmission containment strategies. We believe these model types offer more flexibility to incorporate any type of transition time distributions, hence improving the model's reliability for SARS-CoV-2 and other pathogens. 
\ifCLASSOPTIONcompsoc
  \section*{Acknowledgments}
\else
  \section*{Acknowledgment}
\fi

This material is based on work supported by the National Science Foundation under Grant No.2027336 IIS.

\ifCLASSOPTIONcaptionsoff
  \newpage
\fi

\bibliographystyle{IEEEtran}
\bibliography{Refrences}

\begin{thebibliography}{10}
\providecommand{\url}[1]{#1}
\csname url@samestyle\endcsname
\providecommand{\newblock}{\relax}
\providecommand{\bibinfo}[2]{#2}
\providecommand{\BIBentrySTDinterwordspacing}{\spaceskip=0pt\relax}
\providecommand{\BIBentryALTinterwordstretchfactor}{4}
\providecommand{\BIBentryALTinterwordspacing}{\spaceskip=\fontdimen2\font plus
\BIBentryALTinterwordstretchfactor\fontdimen3\font minus
  \fontdimen4\font\relax}
\providecommand{\BIBforeignlanguage}[2]{{%
\expandafter\ifx\csname l@#1\endcsname\relax
\typeout{** WARNING: IEEEtran.bst: No hyphenation pattern has been}%
\typeout{** loaded for the language `#1'. Using the pattern for}%
\typeout{** the default language instead.}%
\else
\language=\csname l@#1\endcsname
\fi
#2}}
\providecommand{\BIBdecl}{\relax}
\BIBdecl

\bibitem{chowdhury2020covid}
M.~J.~M. Chowdhury, M.~S. Ferdous, K.~Biswas, N.~Chowdhury, and
  V.~Muthukkumarasamy, ``Covid-19 contact tracing: challenges and future
  directions,'' \emph{IEEE Access}, vol.~8, pp. 225\,703--225\,729, 2020.

\bibitem{kretzschmar2020impact}
M.~E. Kretzschmar, G.~Rozhnova, M.~C. Bootsma, M.~van Boven, J.~H. van~de
  Wijgert, and M.~J. Bonten, ``Impact of delays on effectiveness of contact
  tracing strategies for covid-19: a modelling study,'' \emph{The Lancet Public
  Health}, vol.~5, no.~8, pp. e452--e459, 2020.

\bibitem{barrat2020effect}
A.~Barrat, C.~Cattuto, M.~Kivel{\"a}, S.~Lehmann, and J.~Saram{\"a}ki, ``Effect
  of manual and digital contact tracing on covid-19 outbreaks: a study on
  empirical contact data,'' \emph{Journal of the Royal Society Interface},
  vol.~18, no. 178, p. 20201000, 2020.

\bibitem{keeling2020efficacy}
M.~J. Keeling, T.~D. Hollingsworth, and J.~M. Read, ``Efficacy of contact
  tracing for the containment of the 2019 novel coronavirus (covid-19),''
  \emph{J Epidemiol Community Health}, vol.~74, no.~10, pp. 861--866, 2020.

\bibitem{kucharski2020effectiveness}
A.~J. Kucharski, P.~Klepac, A.~J. Conlan, S.~M. Kissler, M.~L. Tang, H.~Fry,
  J.~R. Gog, W.~J. Edmunds, J.~C. Emery, G.~Medley \emph{et~al.},
  ``Effectiveness of isolation, testing, contact tracing, and physical
  distancing on reducing transmission of sars-cov-2 in different settings: a
  mathematical modelling study,'' \emph{The Lancet Infectious Diseases},
  vol.~20, no.~10, pp. 1151--1160, 2020.

\bibitem{cheng2020contact}
H.-Y. Cheng, S.-W. Jian, D.-P. Liu, T.-C. Ng, W.-T. Huang, H.-H. Lin
  \emph{et~al.}, ``Contact tracing assessment of covid-19 transmission dynamics
  in taiwan and risk at different exposure periods before and after symptom
  onset,'' \emph{JAMA internal medicine}, vol. 180, no.~9, pp. 1156--1163,
  2020.

\bibitem{moon2021contact}
S.~A. Moon and C.~M. Scoglio, ``Contact tracing evaluation for covid-19
  transmission in the different movement levels of a rural college town in the
  usa,'' \emph{Scientific reports}, vol.~11, no.~1, pp. 1--12, 2021.

\bibitem{kwok2019epidemic}
K.~O. Kwok, A.~Tang, V.~W. Wei, W.~H. Park, E.~K. Yeoh, and S.~Riley,
  ``Epidemic models of contact tracing: systematic review of transmission
  studies of severe acute respiratory syndrome and middle east respiratory
  syndrome,'' \emph{Computational and structural biotechnology journal},
  vol.~17, pp. 186--194, 2019.

\bibitem{sanche2020high}
S.~Sanche, Y.~T. Lin, C.~Xu, E.~Romero-Severson, N.~Hengartner, and R.~Ke,
  ``High contagiousness and rapid spread of severe acute respiratory syndrome
  coronavirus 2,'' \emph{Emerging infectious diseases}, vol.~26, no.~7, p.
  1470, 2020.

\bibitem{backer2020incubation}
J.~A. Backer, D.~Klinkenberg, and J.~Wallinga, ``Incubation period of 2019
  novel coronavirus (2019-ncov) infections among travellers from wuhan, china,
  20--28 january 2020,'' \emph{Eurosurveillance}, vol.~25, no.~5, p. 2000062,
  2020.

\bibitem{nowzari2015general}
C.~Nowzari, M.~Ogura, V.~M. Preciado, and G.~J. Pappas, ``A general class of
  spreading processes with non-markovian dynamics,'' in \emph{2015 54th IEEE
  Conference on Decision and Control (CDC)}.\hskip 1em plus 0.5em minus
  0.4em\relax IEEE, 2015, pp. 5073--5078.

\bibitem{sherborne2018mean}
N.~Sherborne, J.~C. Miller, K.~B. Blyuss, and I.~Z. Kiss, ``Mean-field models
  for non-markovian epidemics on networks,'' \emph{Journal of mathematical
  biology}, vol.~76, no.~3, pp. 755--778, 2018.

\bibitem{pellis2015exact}
L.~Pellis, T.~House, and M.~J. Keeling, ``Exact and approximate moment closures
  for non-markovian network epidemics,'' \emph{Journal of theoretical biology},
  vol. 382, pp. 160--177, 2015.

\bibitem{boguna2014simulating}
M.~Bogun{\'a}, L.~F. Lafuerza, R.~Toral, and M.~{\'A}. Serrano, ``Simulating
  non-markovian stochastic processes,'' \emph{Physical Review E}, vol.~90,
  no.~4, p. 042108, 2014.

\bibitem{van2011n}
P.~Van~Mieghem, ``The n-intertwined sis epidemic network model,''
  \emph{Computing}, vol.~93, no. 2-4, pp. 147--169, 2011.

\bibitem{sahneh2013generalized}
F.~D. Sahneh, C.~Scoglio, and P.~Van~Mieghem, ``Generalized epidemic mean-field
  model for spreading processes over multilayer complex networks,''
  \emph{IEEE/ACM Transactions on Networking}, vol.~21, no.~5, pp. 1609--1620,
  2013.

\bibitem{pastor2015epidemic}
R.~Pastor-Satorras, C.~Castellano, P.~Van~Mieghem, and A.~Vespignani,
  ``Epidemic processes in complex networks,'' \emph{Reviews of modern physics},
  vol.~87, no.~3, p. 925, 2015.

\bibitem{newman2002spread}
M.~E. Newman, ``Spread of epidemic disease on networks,'' \emph{Physical review
  E}, vol.~66, no.~1, p. 016128, 2002.

\bibitem{boguna2003absence}
M.~Bogun{\'a}, R.~Pastor-Satorras, and A.~Vespignani, ``Absence of epidemic
  threshold in scale-free networks with degree correlations,'' \emph{Physical
  review letters}, vol.~90, no.~2, p. 028701, 2003.

\bibitem{chakrabarti2008epidemic}
D.~Chakrabarti, Y.~Wang, C.~Wang, J.~Leskovec, and C.~Faloutsos, ``Epidemic
  thresholds in real networks,'' \emph{ACM Transactions on Information and
  System Security (TISSEC)}, vol.~10, no.~4, p.~1, 2008.

\bibitem{goltsev2012localization}
A.~V. Goltsev, S.~N. Dorogovtsev, J.~G. Oliveira, and J.~F. Mendes,
  ``Localization and spreading of diseases in complex networks,''
  \emph{Physical review letters}, vol. 109, no.~12, p. 128702, 2012.

\bibitem{diekmann2010construction}
O.~Diekmann, J.~Heesterbeek, and M.~G. Roberts, ``The construction of
  next-generation matrices for compartmental epidemic models,'' \emph{Journal
  of the Royal Society Interface}, vol.~7, no.~47, pp. 873--885, 2010.

\bibitem{diekmann2000mathematical}
O.~Diekmann and J.~A.~P. Heesterbeek, \emph{Mathematical epidemiology of
  infectious diseases: model building, analysis and interpretation}.\hskip 1em
  plus 0.5em minus 0.4em\relax John Wiley \& Sons, 2000, vol.~5.

\bibitem{newman2006finding}
M.~E. Newman, ``Finding community structure in networks using the eigenvectors
  of matrices,'' \emph{Physical review E}, vol.~74, no.~3, p. 036104, 2006.

\bibitem{johansson2021sars}
M.~A. Johansson, T.~M. Quandelacy, S.~Kada, P.~V. Prasad, M.~Steele, J.~T.
  Brooks, R.~B. Slayton, M.~Biggerstaff, and J.~C. Butler, ``Sars-cov-2
  transmission from people without covid-19 symptoms,'' \emph{JAMA network
  open}, vol.~4, no.~1, pp. e2\,035\,057--e2\,035\,057, 2021.

\bibitem{ren2021evidence}
X.~Ren, Y.~Li, X.~Yang, Z.~Li, J.~Cui, A.~Zhu, H.~Zhao, J.~Yu, T.~Nie, M.~Ren
  \emph{et~al.}, ``Evidence for pre-symptomatic transmission of coronavirus
  disease 2019 (covid-19) in china,'' \emph{Influenza and other respiratory
  viruses}, vol.~15, no.~1, pp. 19--26, 2021.

\bibitem{cevik2020sars}
M.~Cevik, M.~Tate, O.~Lloyd, A.~E. Maraolo, J.~Schafers, and A.~Ho,
  ``Sars-cov-2, sars-cov, and mers-cov viral load dynamics, duration of viral
  shedding, and infectiousness: a systematic review and meta-analysis,''
  \emph{The Lancet Microbe}, 2020.

\end{thebibliography}

%




\end{document}